\documentclass[10pt,journal,letterpaper]{IEEEtran}

\usepackage{ifpdf}
\ifpdf
   \usepackage[pdftex]{graphicx}
  \usepackage{mathrsfs}
  \usepackage{amsmath}
  \usepackage[square, comma, sort&compress, numbers]{natbib}
  \usepackage{algorithm}
  \usepackage{algorithmicx}
  \usepackage{algpseudocode}

  \usepackage{enumerate}
  \usepackage{multirow}

  \usepackage{amsthm}
  \usepackage{color}
  \usepackage[square,numbers,sort&compress]{natbib}

  \usepackage{array}
  \newcommand{\PreserveBackslash}[1]{\let\temp=\\#1\let\\=\temp}
  \newcolumntype{C}[1]{>{\PreserveBackslash\centering}p{#1}}
  \newcolumntype{R}[1]{>{\PreserveBackslash\raggedleft}p{#1}}
  \newcolumntype{L}[1]{>{\PreserveBackslash\raggedright}p{#1}}

  \usepackage{marvosym}
  \usepackage{CJK}
  \usepackage{amsfonts}
  \usepackage{subfigure}

  \usepackage{multirow}

  \newtheorem{myDef}{Definition}

\else
\fi

\ifCLASSINFOpdf
\else
\fi

\begin{document}
%

\title{Cloud-Based Approximate Constrained Shortest Distance Queries over Encrypted Graphs with Privacy Protection}

\author{Meng Shen,~\IEEEmembership{Member,~IEEE,}
        Baoli Ma,
        Liehuang Zhu,~\IEEEmembership{Member,~IEEE,}
        Rashid Mijumbi,~\IEEEmembership{Member,~IEEE,}\\
        Xiaojiang Du,~\IEEEmembership{Senior Member,~IEEE,}
        and Jiankun Hu,~\IEEEmembership{Senior Member,~IEEE}
\IEEEcompsocitemizethanks{
\IEEEcompsocthanksitem This work was supported in part by the National Science Foundation of China under Grant 61602039,
in part by the Beijing Natural Science Foundation under Grant 4164098, and in part by the China National Key Research and Development Program under Grant 2016YFB0800301.
\IEEEcompsocthanksitem M. Shen, B. Ma, and L. Zhu are with Beijing Engineering Research Center of High Volume Language Information Processing and Cloud Computing Applications, School of Computer Science, Beijing Institute of Technology, Beijing, China. Email: \{shenmeng, baolimasmile, liehuangz\}@bit.edu.cn. Prof. Liehuang Zhu is the corresponding author.
\IEEEcompsocthanksitem R. Mijumbi is with the Bell Labs CTO, Nokia, Dublin, Ireland. Email: rashid.mijumbi@nokia.com.
\IEEEcompsocthanksitem X. Du is with the Department of Computer and Information Sciences, Temple University, Philadelphia, USA. Email: dxj@ieee.org.
\IEEEcompsocthanksitem J. Hu is with the School of Engineering and IT, University of New South Wales (UNSW), Canberra, Australia. Email: J.Hu@adfa.edu.au.
}
}

\maketitle

\begin{abstract}
Constrained shortest distance (CSD) querying is one of the fundamental graph query primitives, which finds the shortest distance from an origin to a destination in a graph with a constraint that the total cost does not exceed a given threshold.
CSD querying has a wide range of applications, such as routing in telecommunications and transportation.
With an increasing prevalence of cloud computing paradigm,
graph owners desire to outsource their graphs to cloud servers.
In order to protect sensitive information, these graphs are usually encrypted before being outsourced to the cloud.
This, however, imposes a great challenge to CSD querying over encrypted graphs.
Since performing constraint filtering is an intractable task, existing work mainly focuses on unconstrained shortest distance queries. CSD querying over encrypted graphs remains an open research problem.

In this paper, we propose \texttt{Connor}, a novel graph encryption scheme that enables approximate CSD querying. \texttt{Connor} is built based on an efficient, tree-based ciphertext comparison protocol, and makes use of symmetric-key primitives and the somewhat homomorphic encryption,
making it computationally efficient.
Using \texttt{Connor}, a graph owner can first encrypt privacy-sensitive graphs and then outsource them to the cloud server, achieving the necessary privacy without losing the ability of querying. Extensive experiments with real-world datasets demonstrate the effectiveness and efficiency of the proposed graph encryption scheme.

\end{abstract}


\IEEEpeerreviewmaketitle

\section{Introduction}

\IEEEPARstart{R}ecent years have witnessed the prosperity of applications based on graph-structured data \cite{Graph:Encryption:for:Approximate:Shortest:Distance:Queries,
Effective:Indexing:Approximate:Constrained:Shortest:Path:Queries},
such as online social networks, road networks,
web graphs \cite{Shen2017Classification}, biological networks, and communication networks \cite{Shen2014Towards, Xu2015Achieving}.
Consequently, many systems for managing, querying, and analyzing massive graphs have been proposed in both academia (e.g., GraphLab \cite{GraphLab:A:New:Framework:For:Parallel:Machine:Learning},
Pregel \cite{Pregel:a:system:for:large:scale:graph:processing}
and TurboGraph \cite{TurboGraph})
and industry (e.g., Titan, DEX and GraphBase).
With the prevalence of cloud computing,
graph owners (e.g., enterprises and startups for graph-based services) desire to outsource their graph databases to a cloud server, which raises a great concern regarding privacy.
An intuitive way to enhance data privacy is encrypting graphs before outsourcing them to the cloud. This, however, usually comes at the price of inefficiency, because it is quite difficult to perform operations over encrypted graphs.

Shortest distance querying is one of the most fundamental graph operations,
which finds the shortest distance, according to a specific criterion, for a given pair of source and destination in a graph.
In practice, however, users may consider multiple criteria when performing shortest distance queries \cite{Effective:Indexing:Approximate:Constrained:Shortest:Path:Queries}.
Taking the road network as an example,
a user may want to know the shortest distance, in terms of travelling time, between two cities within a budget for total toll payment.
This problem can be represented by a constrained shortest distance (CSD) query, which finds the shortest distance based on one criterion with one or more constraints on other criteria.

In this paper, we focus on single-constraint CSD queries.
This is because most practical problems can be represented as a single-constraint CSD query.
For instance, such a query on a communication network could
return the minimum cost from a starting node to a terminus node, with a threshold on routing delay.
In addition, multi-constraint CSD queries can usually be decomposed into a group of sub-queries, each of which can be abstracted as a single-constraint CSD query.
Formally, a CSD query\footnote{For simplicity, we refer to single-constraint CSD queries as CSD queries hereafter.} is such that:
given an origin $s$, a destination $t$, and a cost constraint $\theta$, finding the shortest distance between $s$ and $t$ whose total cost $c$ does not exceed $\theta$.

Existing studies in this area can be roughly classified into two categories.
The \emph{first} category mainly focuses on the CSD query problem over unencrypted graphs \cite{Bicriterion:path:problems,
Approximation:schemes:for:restricted:shortest:path:problem,
Multiobjective:optimization:Improved:FPTAS:shortest:paths:and:non-linear:objectives:with:applications,
Route:Planning:Bicycles-Exact:Constrained:Shortest:Paths:Made:Practical:via:Contraction:Hierarchy,
Effective:Indexing:Approximate:Constrained:Shortest:Path:Queries}.
However, these methods cannot be easily applied in the encrypted graph environment, because many operations on plain graphs required in these methods (e.g., addition, multiplication, and comparison) cannot be carried out successfully without a special design for encrypted graphs.
The \emph{second} category aims at enabling the shortest distance (or shortest path) queries over encrypted graphs \cite{Graph:Encryption:for:Approximate:Shortest:Distance:Queries,
Shortest:Paths:and:Distances:with:Differential:Privacy}.
They usually adopt distance oracles such that the approximate distance between any two vertices can be efficiently computed, e.g., in a sublinear way.
The main limitation of these approaches is that they are incapable of performing constraint filtering over the cloud-based encrypted graphs.
Therefore, they cannot be directly applied to answering CSD queries.

Motivated by the limitations of existing schemes,
our goal in this paper is to design a practical graph encryption scheme that enables CSD queries over encrypted graphs.
As the CSD problem over plain graphs has been proved to be NP-hard \cite{Approximation:schemes:for:restricted:shortest:path:problem},
existing studies (e.g., \cite{Effective:Indexing:Approximate:Constrained:Shortest:Path:Queries}) usually resort to approximate solutions, which guarantee that the resulting distance is no longer than $\alpha$ times of the shortest distance (where $\alpha$ is an approximation ratio predefined by graph owners),
subject to the cost constraint $\theta$.
The encryption of graphs would make the CSD problem even more complicated. Hence, we also focus on devising an approximate solution.

Specifically, this paper presents \texttt{Connor}, a novel graph encryption scheme targeting the approximate CSD querying
over encrypted graphs.
\texttt{Connor} is built on a secure 2-hop cover labeling index (2HCLI),
which is a type of distance oracle such that the approximate distance between any two vertices in a graph can be efficiently computed \cite{Graph:Encryption:for:Approximate:Shortest:Distance:Queries, Effective:Indexing:Approximate:Constrained:Shortest:Path:Queries}.
The vertices of the graph in the secure 2HCLI are encrypted by particular pseudo-random functions (PRFs).
In order to protect real values of graph attributes while allowing for cost filtering,
we encrypt \emph{costs} and \emph{distances} (between pairs of vertices) by the order-revealing encryption (ORE) \cite{Order:Revealing:Encryption:New:Constructions:Applications:and:Lower:Bounds,
Practical:Order:Revealing:Encryption:with:Limited:Leakage}
and the somewhat homomorphic encryption (SWHE) \cite{boneh2005evaluating}, respectively.
Based on the ORE,
we design a simple but efficient tree-based ciphertexts comparison protocol, which can accelerate the constraint filtering process on the cloud side.

The main contributions of this paper are as follows.
\begin{enumerate}
  \item We propose a novel graph encryption scheme, \texttt{Connor}, which enables the approximate CSD querying. It can answer an $\alpha$-CSD query in milliseconds and thereby achieves computational efficiency.

  \item We design a tree-based ciphertexts comparison protocol, which helps us to determine the relationship of the sum of two integers and another integer over their ciphertexts with controlled disclosure. This protocol can also serve as a building block in other relevant application scenarios.

  \item We present a thorough security analysis of \texttt{Connor} and demonstrate that it achieves the latest security definition named CQA2-security \cite{Structured:encryption}.
      We also implement a prototype and conduct extensive experiments on real-world datasets. The evaluation results show the effectiveness and efficiency of the proposed scheme.
\end{enumerate}

To the best of our knowledge,
this is the first work that enables the approximate CSD querying over encrypted graphs.

The rest of this paper is organized as follows.
We summarize the related work in Section \ref{sec:related_work}
and describe the background of the approximate CSD querying
in Section \ref{sec:background}.
We formally define the privacy-preserving approximate CSD
querying problem in Section \ref{sec:PROBLEM DESCRUPTION}.
After that, the construction of \texttt{Connor} is presented in Section \ref{sec:main scheme},
with a detailed description of the tree-based ciphertexts comparison protocol in Section \ref{sec:Tree:Based:Ciphertexts:Comparison:Approach}.
We exhibit the complexity and security analyses in Section \ref{sec:security}, evaluate the proposed scheme through extensive experiments in Section \ref{sec:evaluation},
and conclude this paper in Section \ref{sec:conclusion}.

\section{Related Work}\label{sec:related_work}
In an era of cloud computing, security and privacy become great concerns of cloud service users
\cite{keyword_hu, du_infocom14, Cheng2017A, Wu2014MobiFish,
Wu2014Security, Du2007An}.
Here we briefly summarize the related work
from two aspects, i.e., CSD querying over plain graphs and graph privacy protection.

\textbf{Plain CSD queries.}
The constrained shortest distance/path querying over plain graphs has attracted many research attentions.
Hansen \cite{Bicriterion:path:problems} proposed an augmented Dijkstra's algorithm for exact constrained shortest path
queries without an index.
This method, however, resulted in a significant computational burden.
In order to improve the querying efficiency,
another solution \cite{Multiobjective:optimization:Improved:FPTAS:shortest:paths:and:non-linear:objectives:with:applications}
focused on approximate constrained shortest path queries, which were also index-free.

The state-of-the-art solution to the exact constrained shortest path querying with an index was proposed by Storandt \cite{Route:Planning:Bicycles-Exact:Constrained:Shortest:Paths:Made:Practical:via:Contraction:Hierarchy},
which accelerated query procedure with an indexing technique called contraction hierarchies.
This approach still results in impractically high query processing cost.
Wang et al. \cite{Effective:Indexing:Approximate:Constrained:Shortest:Path:Queries}
proposed a solution to the approximate constrained shortest path querying over large-scale road networks.
This method took full advantage of overlay graph techniques
to construct an overlay graph based on the original graph,
whose size was much smaller than that of the original one.
Consequently, they built a constrained labeling index structure over the overlay graph,
which greatly reduced the query cost.
Unfortunately, all these solutions are merely suitable to perform queries over unencrypted graphs.

\textbf{Graph privacy protection.}
Increasing concerns about graph privacy have been raised with the wide adoption of the cloud computing paradigm over the past decade.
Chase and Kamara \cite{Structured:encryption} first introduced the notion of graph encryption,
where they proposed several constructions for graph operations,
such as adjacency queries and neighboring queries.
Cao et al. \cite{Privacy:Preserving:Query:over:Encrypted:Graph:Structured:Data:in:Cloud:Computing}
defined and solved the problem of privacy-preserving query over encrypted graph data in cloud computing
by utilizing the principle of ``filtering-and-verification''.
They built the feature-based index of a graph in advance and then chose the efficient inner product to carry out the filtering procedure.
Some approaches \cite{Analyzing:graphs:with:node:differential:privacy,
Mining:Frequent:Graph:Patterns:with:Differential:Privacy,
Shortest:Paths:and:Distances:with:Differential:Privacy}
utilized the differential privacy technique to query graphs privately,
which might suffer from weak security.
These studies, however, introduced prohibitively great storage costs and were not practical for large-scale graphs.
Meng et al. \cite{Graph:Encryption:for:Approximate:Shortest:Distance:Queries}
proposed three computationally efficient constructions
that supported the approximate shortest distance querying with distance oracles, which were provably secure against a semi-honest cloud server.

Secure multi-party computation (SMC) techniques
have been widely applied to address the privacy-preserving shortest path problem \cite{Blanton2013Data,
Aly2013Securely,
SMC_SP_Keller2014Efficient,
SMC_Gupta2012A},
as well as other secure computation problems
\cite{Bayatbabolghani2017Secure}.
Aly et al. \cite{Aly2013Securely} focused on the shortest path problem over traditional combinatorial graph in a general multi-party computation setting,
and proposed two protocols for securely computing shortest paths in the graphs.
Blanton et al. \cite{Blanton2013Data} designed data-oblivious algorithms to securely solve the single-source single-destination shortest path problem,
which achieved the optimal or near-optimal performance on dense graphs.
Keller and Scholl \cite{SMC_SP_Keller2014Efficient} designed several oblivious data structures (e.g., priority queues) for SMC and utilized them to compute shortest paths
on general graphs.
Gupta et al. \cite{SMC_Gupta2012A} proposed an SMC-based approach for finding policy-compliant paths that have the least routing cost or satisfy bandwidth demands among different network domains.
However, existing general-purpose SMC solutions for the shortest path problem may result in heavy communication overhead.

Although there are respectable studies on graph querying over encrypted graphs,
the privacy-preserving CSD query remains unsolved.
In this paper, we propose a novel and efficient graph encryption scheme for CSD queries.

\section{Background}\label{sec:background}
This section presents the formal definition of the CSD query problem and introduces the 2HCLI structure for graph queries.

\begin{table}[t]
\renewcommand{\arraystretch}{1.3}
\caption{List of notations} \label{tab:notations}
\centering
\footnotesize{
\begin{tabular}{|L{2.0cm}|L{5.6cm}|}
\hline
\textbf{Notation}  & \textbf{Meaning}  \\ \hline
$G=(V,E)$ & Input graph \\ \hline
$n, m$ & Number of vertices and edges in $G$ \\ \hline
$d(e), c(e)$ & Distance and cost of an edge $e$ \\ \hline
$d(u,v), c(u,v)$ & Distance and cost of the edge from $u$ to $v$ \\ \hline
$s, t, \alpha, \phi, \theta$ & Origin, destination, approximation ratio, amplification factor and cost constraint in an $\alpha$-CSD query \\ \hline
$\Delta, \widetilde{\Delta}$ & Plain and encrypted graph index \\ \hline
$\Delta_{in}(v), \Delta_{out}(v)$ & In- and out-label set associated with vertex $v$ \\ \hline
$d_{\theta}$ & Depth of a cost constraint tree \\ \hline
$\beta$ & Length of a path code \\ \hline
$E(m)$ & ORE ciphertext of $m$ \\ \hline
$\lambda$ & Security parameter \\ \hline
$k$ & Output length of ORE encryption \\ \hline
$z$ & Input length of symmetric encryption algorithms \\ \hline
$\tau_{s,t}$ & Query token \\ \hline
$Y$ & Candidate sets as the outputs of the cost constraint filtering  \\ \hline
$\mathcal{B}$ & Maximum distance over all the sketches \\ \hline
\end{tabular}}
\end{table}

\subsection{Approximate CSD Query}\label{sec:definitions}
Let $G=(V,E)$ be a directed graph\footnote{We refer to $G$ as a directed graph in this paper, unless otherwise specified.} with a vertex set $V$ and an edge set $E$.
Each edge $e \in E$ is associated with a \emph{distance} $d(e) \ge 0$ and a \emph{cost} $c(e) \ge 0$.
We regard the cost $c(e)$ as the constraint.
We denote the set of edges that connect two vertices as a \emph{path}. For a path $P = (e_1, e_2, \dots , e_k)$,
its distance $d(P)$ is defined as $d(P)= \sum_{i=1}^{k} d(e_i)$,
which indicates the distance from its origin to its destination.
Similarly, we define the cost of $P$ as $c(P)= \sum_{i=1}^{k} c(e_i)$.
The notations throughout the paper are summarized in Table \ref{tab:notations}.

Given a graph $G$, an origin vertex $s \in V$, a destination vertex $t \in V$,
and a cost constraint $\theta$, a CSD query
is to find the the shortest distance $d$ between $s$ and $t$ with the total cost no more than $\theta$.
Since the CSD query problem has been proved to be NP-hard
\cite{Approximation:schemes:for:restricted:shortest:path:problem},
we keep in line with existing solutions \cite{Effective:Indexing:Approximate:Constrained:Shortest:Path:Queries}
and focus on proposing an approximate CSD solution in this paper.

Inspired by a common definition of the approximate shortest path query over plain graphs \cite{Effective:Indexing:Approximate:Constrained:Shortest:Path:Queries},
we define the approximate CSD query (i.e., $\alpha$-CSD query) as follows.
\begin{myDef}
($\alpha$-CSD QUERY).
\emph{
Given an origin $s$, a destination $t$, a cost constraint $\theta$ and an approximation ratio $\alpha$,
an $\alpha$-CSD query returns the distance $d(P)$ of a path $P$, such that $c(P) \le \theta$ and $d(P) \le \alpha \cdot d_{opt}$, where $d_{opt}$ is the optimal answer to the exact CSD query with the origin $s$, destination $t$ and cost constraint $\theta$.
}
\end{myDef}

\begin{figure}[t]
\begin{minipage}{0.49\textwidth}
  \centering
  \includegraphics[height=3cm]{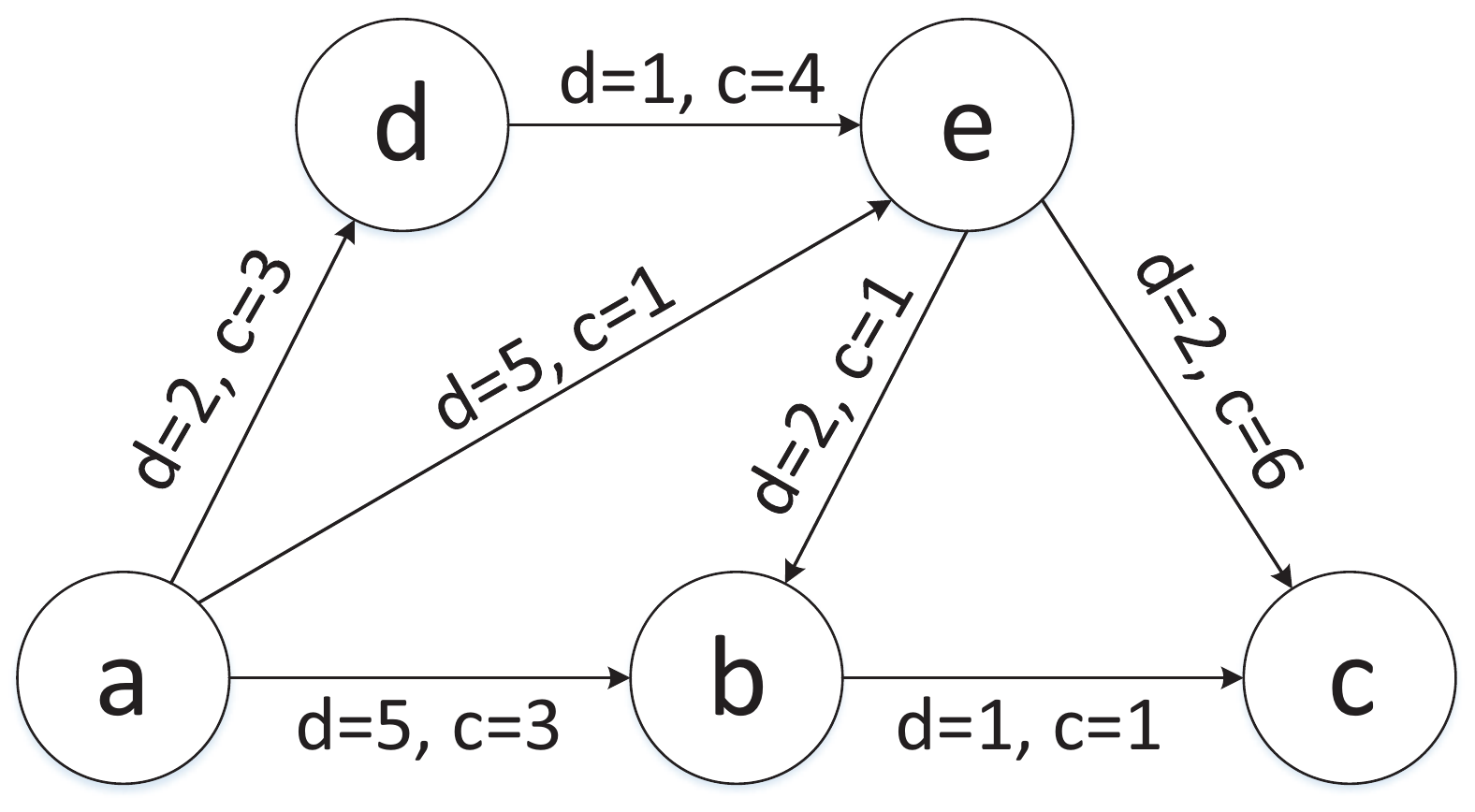}
\end{minipage}
\caption{An example illustrating the $\alpha$-CSD query over a graph.}\label{fig:graph}
\end{figure}

Fig. \ref{fig:graph} shows a simple graph with five vertices,
where the distance and cost of each edge are marked alongside it.
Given an origin $a$, a destination $c$, a cost constraint $\theta=4$, the exact CSD query returns the optimal distance $d_{opt}=6$, where the corresponding path is $(a,b,c)$.
For an approximation ratio $\alpha=1.5$,
a valid answer to the $\alpha$-CSD query with the same parameters (e.g, the origin $a$, the destination $c$, and $\theta=4$) is 8, with the corresponding path $P_\alpha=(a,e,b,c)$.
That is because $d(P_\alpha)=8 < \alpha \cdot d_{opt}=9$ and
$c(P_\alpha)=3 < \theta$.

Based on the above definition,
given two paths $P_{1}$ and $P_{2}$ with the same origin and destination,
we say that $P_{1}$ $\alpha$-\emph{dominates} $P_{2}$ iff $c(P_{1}) \le c(P_{2})$ and $d(P_{1}) \le \alpha \cdot d(P_{2})$.
With this principle, we can reduce the construction complexity of graph index significantly,
because a great deal of redundant entries in the index can be filtered out.
We will make a further illustration in the following subsection.

\subsection{Constructing Labeling Index} \label{sec:2HCLI}
The encrypted index designed in this paper is mainly constructed based on the well-known 2HCLI, which is a special data structure that supports the shortest distance query efficiently \cite{Reachability:and:Distance:Queries:Via:2:Hop:Labels,
Fast:exact:shortest:path:distance:queries:on:large:networks:by:pruned:landmark:labeling,
Effective:Indexing:Approximate:Constrained:Shortest:Path:Queries}.
Here we briefly describe the basic idea of the 2HCLI,
and illustrate its application in building a constrained labeling index.

Given a graph $G=(V,E)$ with a vertex set $V$ and an edge set $E$, each vertex $v \in V$ is associated with an in-label set and an out-label set, which are denoted by $\Delta_{in}(v)$ and $\Delta_{out}(v)$, respectively.
Each entity in $\Delta_{in}(v)$ corresponds to the shortest distance from a vertex $u \in V$ to $v$.
It implies that $v$ is reachable from $u$ by one or more paths, but is not necessarily a neighbor, or 2-hop neighbour, of $u$.
Similarly, each entity in $\Delta_{out}(v)$ corresponds to the shortest distance from $v$ to another vertex $u$ in $V$.
To answer a shortest distance query from an origin $s$ to a destination $t$,
we first find the common vertices in the labels $\Delta_{out}(s)$ and $\Delta_{in}(t)$,
and then select the shortest distance from $s$ to $t$.
Note that the entities in $\Delta_{in}(v)$ and $\Delta_{out}(v)$ are carefully selected \cite{Fast:exact:shortest:path:distance:queries:on:large:networks:by:pruned:landmark:labeling}
so that the distance of any two vertices $s$ and $t$ can be computed by $\Delta_{out}(s)$ and $\Delta_{in}(t)$.

\begin{figure}[t]
\begin{minipage}{0.49\textwidth}
  \centering
  \includegraphics[height=3cm]{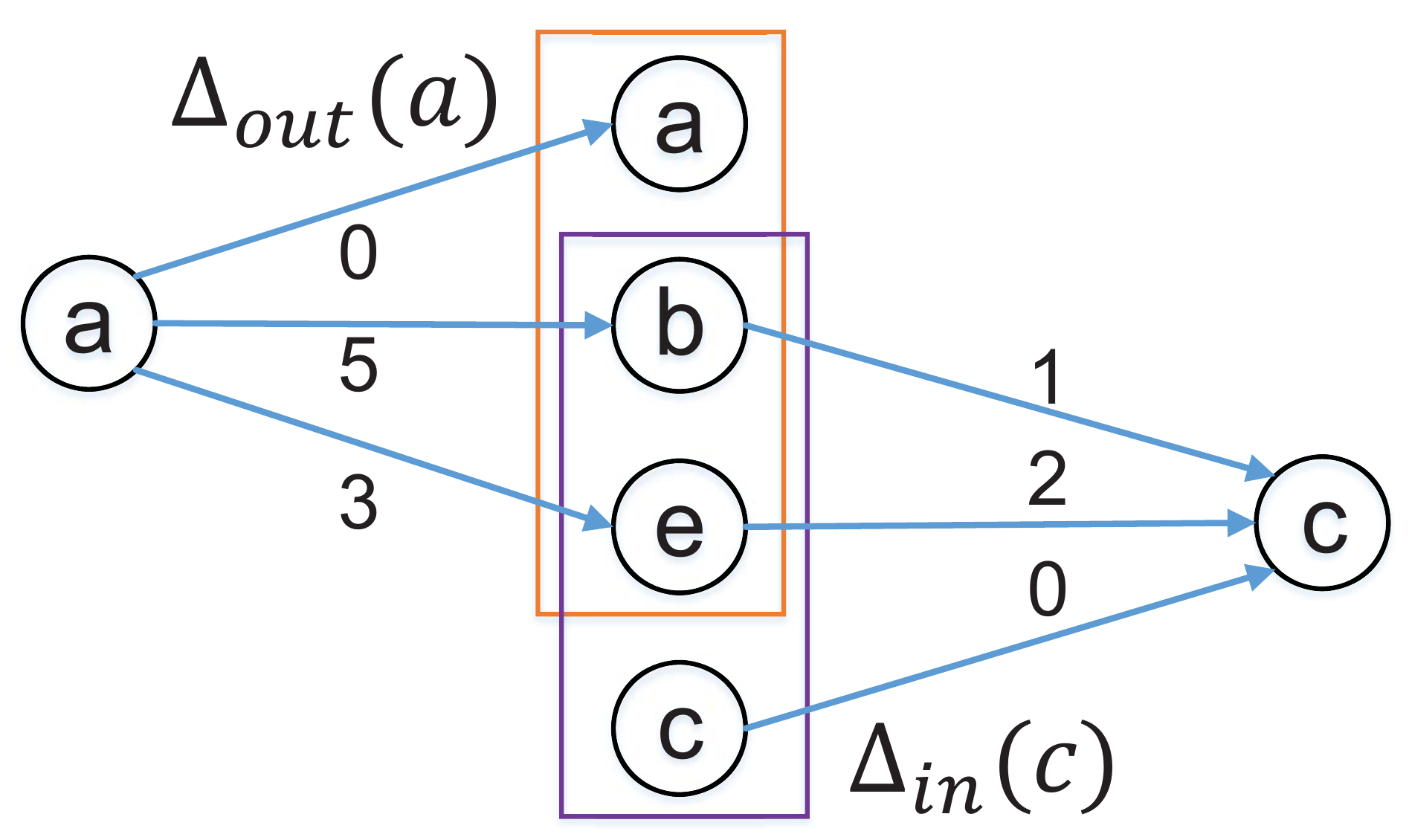}
\end{minipage}
\caption{A 2HCLI example of the basic shortest distance query.
Each entity $d$ in 2HCLI alongside the arrow indicates the shortest distance from the starting vertex to the ending vertex, e.g., the shortest distance from $a$ to $e$ is 3.}\label{fig:2HCLI_1}
\end{figure}

Considering the graph in Fig. \ref{fig:graph},
if we ignore the cost criterion of edges,
the \emph{basic} unconstrained shortest distance query
with an origin $a$ and a destination $c$ can be answered with the help of the 2HCLI, as shown in Fig. \ref{fig:2HCLI_1}.
Given the labels $\Delta_{out}(a)$ and $\Delta_{in}(c)$,
it is easy to obtain the set of common vertices, which consists of vertices $b$ and $e$.
The final answer to the basic shortest distance query should be 5, because $d(a,e)+d(e,c)=5 < d(a,b)+d(b,c)=6$.

\begin{figure}[t]
\begin{minipage}{0.49\textwidth}
  \centering
  \includegraphics[height=3cm]{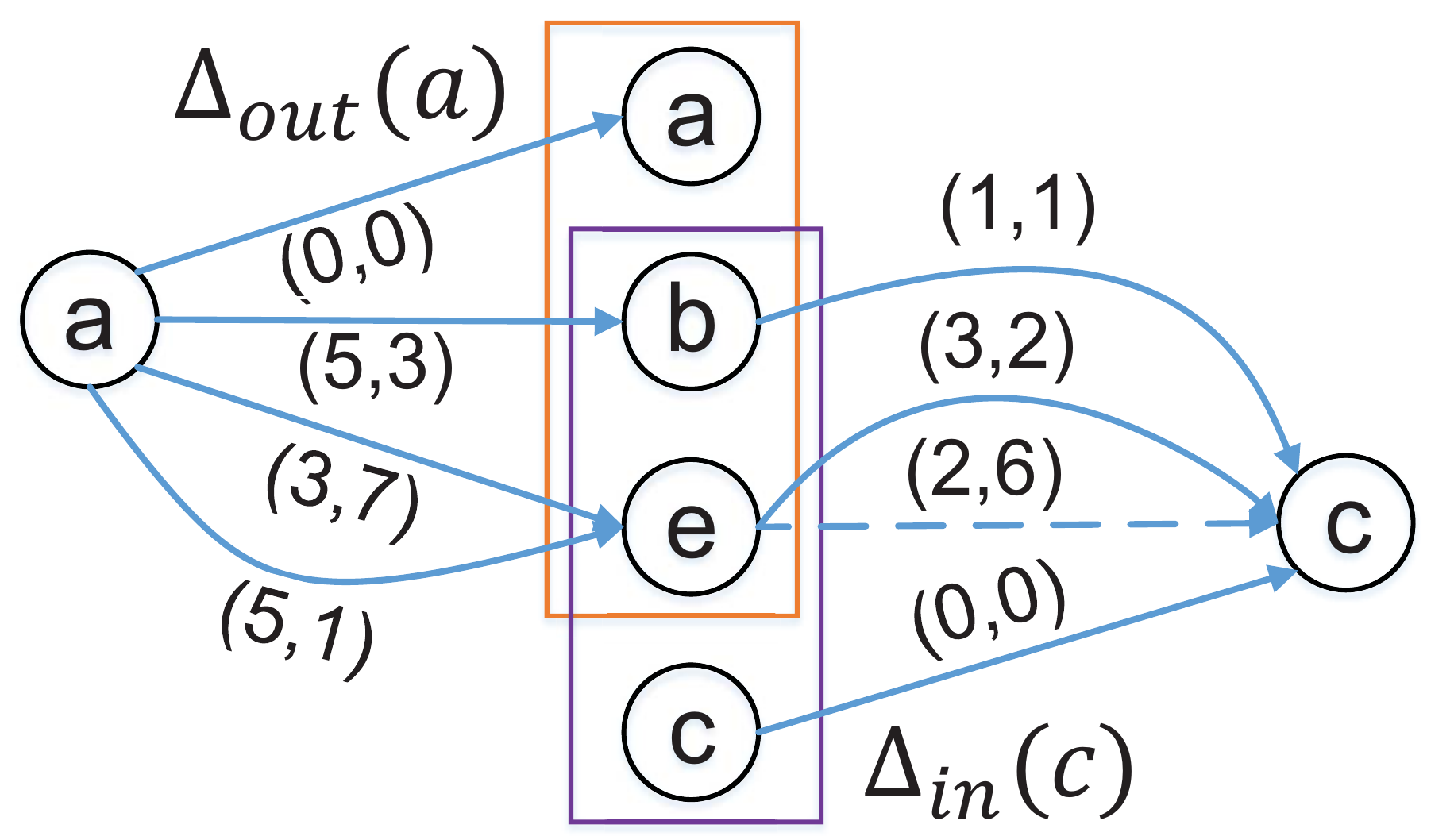}
\end{minipage}
\caption{A 2HCLI example of the exact CSD query.
Each entity $(dis, cost)$ in the 2HCLI alongside the arrow indicates the distance and cost, respectively.
The shortest distance from $a$ to $e$
with a cost constraint $\theta = 4$ is 5.}\label{fig:2HCLI_2}
\end{figure}

Although it is simple and straightforward to construct the 2HCLI for a graph with only the distance criterion,
constructing a labeling index based on the 2HCLI for the CSD query is much more complex.
That is because in the CSD query setting with two types of edge criteria, there might be multiple combinations of distance and cost for each pair of vertices in the labels  $\Delta_{in}(v)$ and $\Delta_{out}(v)$.
For ease of illustration, we also take as an example the graph, as well as the CSD query, in Fig. \ref{fig:graph}.
The corresponding 2HCLI is shown in Fig. \ref{fig:2HCLI_2}, where the 2-tuple alongside each arrow represents the distance and cost from the starting vertex to the ending vertex.
Note that in the shortest distance query in Fig. \ref{fig:2HCLI_1},
the shortest distance from $a$ to $c$ via $e$ is unique.
However, in the CSD query setting depicted in Fig. \ref{fig:2HCLI_2},
there are four possible distances with different costs from $a$ to $c$ via $e$.
Due to the existence of the cost criterion,
the number of possible distances for each pair of vertices could increase dramatically in large-scale graphs,
which results in a higher complexity in constructing the 2HCLI and calculating the answers to a CSD query.

In order to improve the querying efficiency,
we adopt a methodology that combines an \emph{offline} filtering operation and an \emph{online} filtering operation.

The offline filtering aims at reducing the construction complexity of the 2HCLI and decreasing the number of entries in the in-label and out-label sets as many as possible.
We adopt the method proposed in \cite{Effective:Indexing:Approximate:Constrained:Shortest:Path:Queries}.
The entities in the 2HCLI are carefully selected
in such a way that for any CSD query from $u$ to $v$ with a cost constraint $\theta$,
the query can be answered correctly using only the 2HCLI.
Since the construction of the 2HCLI should be independent of the cost constraint in specific CSD queries,
we can use the definition of $\alpha$-\emph{domination} to filter out redundant entries in the in- and out-label sets.

\begin{figure}[t]
\begin{minipage}{0.49\textwidth}
  \centering
  \includegraphics[height=2.6cm]{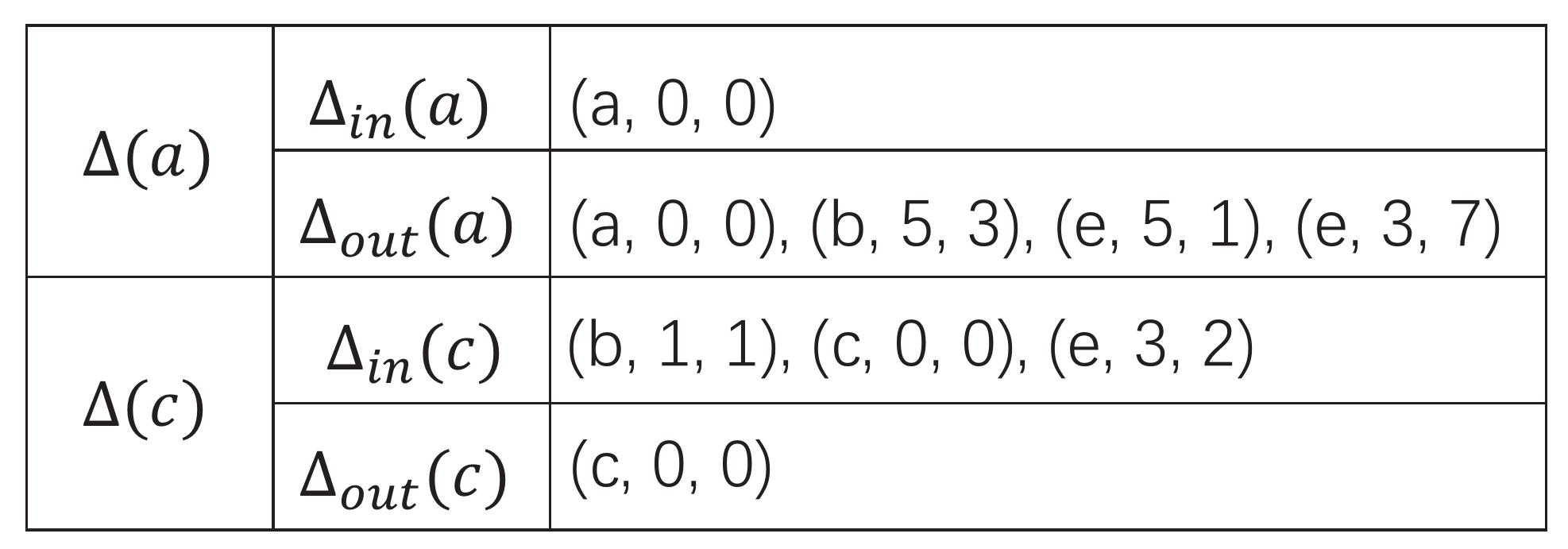}
\end{minipage}
\caption{The resulting 2HCLI after performing the offline filtering on the original 2HCLI in Fig. \ref{fig:2HCLI_2}.
Each entity $(u, d, c)$ in the 2HCLI indicates the vertex identifier, distance and cost, respectively.
The answer to the approximate CSD query (i.e., the origin $a$, the destination $c$, $\alpha=1.5$, and $\theta=4$) is 6, which happens to be the answer to the exact CSD query.}\label{fig:2HCLI}
\end{figure}

Taking for example the two entries from $e$ to $c$ with $\alpha=1.5$ in Fig. \ref{fig:2HCLI_2},
the path $P_{ec}^1 = (e,b,c)$ with the $(dis,cost)$-tuple of (3,2) $\alpha$-\emph{dominates} another path $P_{ec}^2=(e,c)$ with the $(dis,cost)$-tuple of (2,6).
Therefore, the entry corresponding to the path $P_{ec}^2$ can be filtered out (as depicted by a dashed arrow), which helps to reduce the number of entries in $\Delta_{in}(c)$.
The resulting 2HCLI is exhibited in Fig. \ref{fig:2HCLI}.
We refer the reader to \cite{Effective:Indexing:Approximate:Constrained:Shortest:Path:Queries} for more construction details.

The online filtering aims at selecting the possibly valid answers to a given CSD query, based on only the 2HCLI.
For instance, given an $\alpha$-CSD query from $a$ to $c$ with a cost constraint $\theta = 4$,
we can first find the common vertex set $V'$ between $\Delta_{out}(a)$ and $\Delta_{in}(c)$,
and then return the minimum $d(a, v) + d(v, c)$ with $c(a, v) + c(v, c) \le \theta$ for each $v \in V'$.
Since the above comparisons should be conducted with the corresponding ciphertexts,
an efficient online filtering approach will be devised in
Section \ref{sec:Tree:Based:Ciphertexts:Comparison:Approach}.

\section{Problem Formulation}\label{sec:PROBLEM DESCRUPTION}
This section presents the system model and the security model of the privacy-preserving $\alpha$-CSD querying, as well as
the preliminaries of the proposed graph encryption scheme.

\subsection{System Model}
\begin{figure}[t]
\begin{minipage}{0.49\textwidth}
  \centering
  \includegraphics[height=2.5cm]{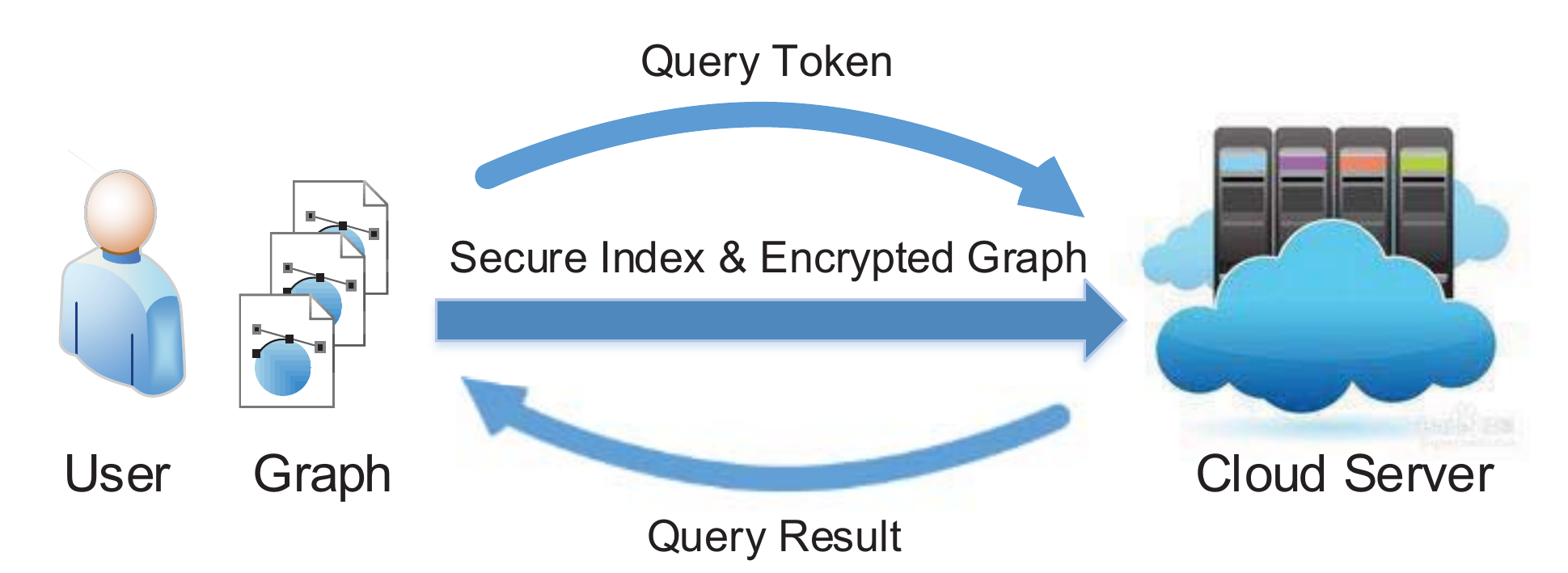}
\end{minipage}
\caption{The system model of privacy-preserving CSD query scheme.}\label{fig:system_model}
\end{figure}

We adopt the general system model in the literature
\cite{Structured:encryption,
Graph:Encryption:for:Approximate:Shortest:Distance:Queries}
for the privacy-preserving $\alpha$-CSD querying,
as illustrated in Fig. \ref{fig:system_model},
which mainly involves two types of entities,
namely a \emph{user} and a \emph{cloud server}.

The \emph{user} constructs the secure searchable index for the graph and outsources the encrypted index along with the encrypted graph to the cloud server.
When the user, say Alice, performs an $\alpha$-CSD query over her encrypted graph,
she first generates a query token
and then submits it to the cloud server.
Upon receiving Alice's query token, the cloud server executes the pre-designed query algorithms to match entries in the secure index with the token.
Finally, the cloud server replies the user with the answer to the
$\alpha$-CSD query.

The graph encryption scheme is formally defined as follows.

\begin{myDef}
(GRAPH ENCRYPTION).
\emph{
A graph encryption scheme $\Pi =(KeyGen, Setup, Query)$ consists of three polynomial-time algorithms that work as follows:}
\begin{itemize}
  \item \textbf{$(K, pk,sk) \gets KeyGen(\lambda)$}\emph{: is a probabilistic secret key generation algorithm that takes as input a security parameter $\lambda$ and outputs a secret key $K$ and a public/secret-key pair $(pk, sk)$.}
  \item \textbf{$\widetilde{\Delta} \gets Setup(\alpha, K, pk, sk, \phi, G)$}\emph{: is a graph encryption algorithm that
  takes as input an approximation ratio $\alpha$, a secret keys $K$, a key pair $(pk, sk)$, an amplification factor $\phi$ and a graph $G$,
  and outputs a secure index $\widetilde{\Delta}$.}
  \item \textbf{$(dist_{q}, \bot) \gets Query((K, pk, sk, \Phi, q), \widetilde{\Delta})$}\emph{: is a two-party protocol between a user that holds a secret key $K$, a key pair $(pk, sk)$ and a query $q$, and a cloud server that holds an encrypted graph index $\widetilde{\Delta}$.
  After executing this protocol,
  the user receives the distance $dist_{q}$ as the query result and the cloud server receives a terminator $\bot$.}
\end{itemize}
\end{myDef}

\subsection{Security Model}
Graph encryption is a generalization of symmetric searchable encryption (SSE) \cite{Practical:techniques:for:searches:on:encrypted:data,
Searchable:Symmetric:Encryption:Improved:Definitions:and:Efficient:Constructions,
Dynamic:searchable:symmetric:encryption,
Dynamic:Searchable:Encryption:in:Very:Large:Databases:Data:Structures:and:Implementation,
Practical:Dynamic:Searchable:Encryption:with:Small:Leakage}.
Thus, we adopt the security definition of SSE settings in our graph encryption scheme.
This security definition is consistent with the latest proposed security definition in \cite{Searchable:Symmetric:Encryption:Improved:Definitions:and:Efficient:Constructions,
2011:Searchable:symmetric:encryption:Improved:definitions:and:efficient:constructions,
Structured:encryption},
which is also known as CQA2-security (i.e., the chosen-query attack security).
Now we present the formal CQA2-security definition as follows.

\begin{myDef}
(CQA2-security model).
\emph{Let $\Pi =(KeyGen, Setup, Query)$ be a graph encryption scheme and consider the following probabilistic experiments where $\mathcal{A}$ is a semi-honest adversary,
$\mathcal{S}$ is a simulator,
and $\mathcal{L}_{Setup}$ and $\mathcal{L}_{Query}$ are (stateful) leakage functions.
}
\end{myDef}

\textbf{Real$_{\Pi, \mathcal{A}}(\lambda)$}:
\begin{itemize}
  \item $\mathcal{A}$ outputs a graph $G$, an approximation ratio $\alpha$ and an amplification factor $\phi$.
  \item The challenger begins by running $Gen(1^{\lambda})$ to generate a secret key $K$ and a public/secret-key pair $(pk, sk)$, and then computes the encrypted index $\widetilde{\Delta}$ by $Setup(\alpha, K, pk, sk, \phi, G)$.
      The challenger sends the encrypted index $\widetilde{\Delta}$ to $\mathcal{A}$.
  \item $\mathcal{A}$ makes a polynomial number of adaptive queries, and for each query $q$,
  $\mathcal{A}$ and the challenger execute $Query((K, pk, sk, \Phi, q), \widetilde{\Delta})$.
  \item $\mathcal{A}$ computes a bit $b \in \{0,1\}$ as the output of the experiment.
\end{itemize}

\textbf{Ideal$_{\Pi, \mathcal{A}, \mathcal{S}}(\lambda)$}:
\begin{itemize}
  \item $\mathcal{A}$ outputs a graph $G$, an approximation ratio $\alpha$ and an amplification factor $\phi$.
  \item Given the leakage function $\mathcal{L}_{Setup}(G)$,
  $\mathcal{S}$ simulates a secure graph index $\widetilde{\Delta}^{*}$ and sends it to $\mathcal{A}$.
  \item $\mathcal{A}$ makes a polynomial number of adaptive queries.
  For each query $q$, $\mathcal{S}$ is given the leakage function $\mathcal{L}_{Query}(G, Q)$,
  and $\mathcal{A}$ and $\mathcal{S}$ execute a simulation of $Query$,
  where $\mathcal{A}$ is playing the role of the cloud server and $\mathcal{S}$ is playing the role of the user.
  \item $\mathcal{A}$ computes a bit $b \in \{0,1\}$ as the output of the experiment.
\end{itemize}

 $\qquad$ We say that the graph encryption
scheme $\Pi =(KeyGen, Setup, Query)$
 is $(\mathcal{L}_{Setup}, \mathcal{L}_{Query})$-secure against the adaptive chosen-query attack,
 if for all PPT adversaries $\mathcal{A}$,
 there exists a PPT simulator $\mathcal{S}$ such that
 \begin{eqnarray*}
|\textbf{Pr}[{\textbf{Real}}_{\Pi, \mathcal{A}}(\lambda) = 1] - \textbf{Pr}[{\textbf{Ideal}}_{\Pi, \mathcal{A}, \mathcal{S}}(\lambda) = 1]| \le negl(\lambda),
\end{eqnarray*}
 where $negl(\lambda)$ is a negligible function.

\subsection{Preliminaries}
Now we briefly introduce an encryption technique employed in our design,
i.e., the order-revealing encryption.

\emph{Order-revealing encryption (ORE)} is a generalization of the order-preserving encryption (OPE) scheme, but provides stronger security guarantees.
As pointed by Naveed et al. \cite{Inference:Attacks:on:Property:Preserving:Encrypted:Databases},
the OPE-encrypted databases are extremely vulnerable to \emph{inference attacks}.
To address this limitation, the ORE scheme has been proposed
\cite{Order:Revealing:Encryption:New:Constructions:Applications:and:Lower:Bounds,
Practical:Order:Revealing:Encryption:with:Limited:Leakage},
which is a tuple of three algorithms
$\Pi=(ORE.Setup, ORE.Encrypt, ORE.Compare)$ described as follows:
\begin{itemize}
  \item ORE.Setup($1^{\lambda}$)$\to sk$: Input a security parameter $\lambda$, output the secret key $sk$.
  \item ORE.Encrypt($sk, m$)$\to ct$: Input a secret key $sk$ and a message $m$, output a ciphertext $ct$.
  \item ORE.Compare($ct_{1}, ct_{2}$)$\to z$: Input two ciphertexts $ct_{1}$ and $ct_{2}$, output a bit $r \in \{0,1\}$,
  which indicates the greater-than or less-than relationship of the corresponding plaintexts $m_{1}$ and $m_{2}$.
\end{itemize}

\section{Construction of \texttt{Connor}}\label{sec:main scheme}
In this section, we introduce our graph encryption scheme \texttt{Connor} for the privacy-preserving $\alpha$-CSD querying.

\subsection{Construction Overview}
The construction process is based on two particular pseudo-random functions $h$ and $g$, and a somewhat homomorphic encryption (SWHE) scheme.
In this paper, we adopt the concrete instantiation of a SWHE scheme in the literature \cite{boneh2005evaluating}.
The parameters of $h$ and $g$ are illustrated in Equation \eqref{eq:parameters},
\begin{subequations} \label{eq:parameters}
\begin{align}
& h: {\{0,1\}}^{\lambda} \times {\{0,1\}}^{*} \to {\{0,1\}}^{\lambda} \label{eq:parameters-1} \\
& g: {\{0,1\}}^{\lambda} \times {\{0,1\}}^{*} \to {\{0,1\}}^{\lambda + z + k} \label{eq:parameters-2}
\end{align}
\end{subequations}
where $\lambda$ is the security parameter, and
$k$ and $z$ are the output lengths of the ORE
and SWHE encryptions, respectively.

We start with a straightforward construction $GraphEnc_{1}=(KeyGen, Setup, Query)$ as follows, including:
\begin{itemize}
  \item \textbf{KeyGen:}
  Given the security parameter $\lambda$,
  the \emph{user} randomly generates a secret key $K$ and a pair of public and secret keys $(pk, sk)$ for SWHE.

  \item \textbf{Setup:}
  Given an original graph $G$, an approximation ratio $\alpha$, and an amplification factor $\phi$,
  the \emph{user} obtains the encrypted graph index by using Algorithm \ref{Algorithm:A straightforward approach:setup}.
  The 2HCLI $\Delta=\{\Delta_{out}, \Delta_{in}\}$ of $G$ can be generated by the method described in Section \ref{sec:2HCLI}.

  Let $\mathcal{B}$ be the maximum distance over all the sketches and $N=2\mathcal{B}+1$.
  Motivated by the literature \cite{Graph:Encryption:for:Approximate:Shortest:Distance:Queries},
  each distance $d_{u,v}$ is encrypted as $2^{N-d_{u,v}}$ by the SWHE to protect its real value (line 8).
  Considering that $2^x +2^y$ is bounded by $2^{max(x,y)-1}$,
  the SWHE encryption of distance allows for obtaining the minimum sum over a certain number of distance pairs.

  Each cost $c_{u,v}$, multiplied by the amplification factor $\phi$, is encrypted by the ORE encryption (line 9).
  $\phi$ is a big integer and should be carefully selected to enlarge the plaintext space of $c_{u,v}$.
  In practice, the product of $\phi$ and the maximum cost value over all the sketches should be sufficiently large (e.g., at least $2^{80}$), which is used to provide a sufficient randomness to the inputs.
  Since $\phi$ is kept private by the \emph{user},
  the \emph{cloud server} cannot learn the real values of $c_{u,v}$.
  \item \textbf{Query:}
  To perform an $\alpha$-CSD query with an origin $s$, a destination $t$, and a cost constraint $\theta$,
  the \emph{user} generates query tokens
  $\tau_{s}=h(K, s||1)$ and $\tau_{t}=h(K, t||2)$,
  and sends them to the \emph{cloud server}.
  The \emph{cloud server} obtains $I_{out}[\tau_{s}]$ and $I_{in}[\tau_{t}]$ from the index.
  For each encrypted vertex identifier $v$ that appears in
  both $I_{out}[\tau_{s}]$ and $I_{in}[\tau_{t}]$,
  the \emph{cloud server} performs a cost constraint filtering operation (which will be described in details in Section \ref{sec:Tree:Based:Ciphertexts:Comparison:Approach}),
  and adds each pair $(D_{s,v}, D_{v,t})$ which satisfies the cost constraint $\phi \theta$ into a candidate set $Y$.
  Note that the cost constraint is multiplied by $\phi$ because we encrypt the cost $\phi c_{u,v}$, instead of $c_{u,v}$.

  Then, the \emph{cloud server} directly obtains $d = \sum_{i=1}^{|Y|}d_{i}$, where $d_{i} = $ SWHE.Eval$(\times, D_{s,v}^{i}, D_{v,t}^{i})$ for each pair $(D_{s,v}^{i}, D_{v,t}^{i})$ in $Y$.
  The correctness of the above calculation follows homomorphic properties of SWHE.
  We refer the readers to \cite{Graph:Encryption:for:Approximate:Shortest:Distance:Queries}
  for more details.

  Finally, the \emph{cloud server} returns $d$ to the \emph{user}, who, in turn, obtains the answer to the $\alpha$-CSD query by decrypting $d$ with its secret key $sk$.
\end{itemize}

Note that this straightforward approach does not only correctly answer the $\alpha$-CSD query over encrypted graphs, but also protects the vertex identifier, distance, and cost information.

However, the encrypted graph index obtained from Algorithm \ref{Algorithm:A straightforward approach:setup},
without performing any queries, still results in information leakage.
On one hand, it reveals the length of each encrypted sketch, i.e., $I_{out}[u]$ and $I_{in}[u]$, as well as the order information of ORE-encrypted costs in all sketches.
On the other hand,
it also discloses the number of common vertices between $I_{out}[u]$ and $I_{in}[v]$, which indicates the number of vertices that connect $u$ to $v$.
In particular, if the \emph{cloud server} knows that there is no common vertex between $I_{out}[u]$ and $I_{in}[v]$,
it learns that $u$ cannot reach $v$.

\renewcommand{\algorithmicrequire}{\textbf{Input:}}
\renewcommand{\algorithmicensure}{\textbf{Output:}}

\begin{algorithm}[t]
  \footnotesize
  \caption{Setup algorithm for $GraphEnc_{1}$}\label{Algorithm:A straightforward approach:setup}
  \begin{algorithmic}[1]
  \Require A secret key $K$, a key pair $(pk, sk)$, an approximation ratio $\alpha$, an amplification factor $\phi$, and an original graph $G$.
  \Ensure The encrypted graph index $\widetilde{\Delta}$.

  \State Generate the \emph{2-hop labeling index} $\Delta=\{\Delta_{out}, \Delta_{in}\}$ from $G$.
  \State Initialize two dictionaries $I_{out}$ and $I_{in}$.
  \State Let $\mathcal{B}$ be the maximum distance over all the sketches and set $N = 2\mathcal{B} +1$.

  \For {each $u \in G$}
        \State Set $T_{out,u}=h(K, u||1)$, $T_{in,u}=h(K, u||2)$.
        \For {each $(v, d_{u,v}, c_{u,v}) \in \Delta_{out}(u)$}
            \State Compute $V = h(K, v||0)$.
            \State Compute $D_{u,v}=$ SWHE.Enc$(pk, 2^{N-d_{u,v}})$.
            \State Compute $C_{u,v}=$ ORE.Enc$(K, \phi  c_{u,v})$.
            \State Insert ($V,D_{u,v}, C_{u,v}$) into the dictionary $I_{out}[T_{out,u}]$.
        \EndFor
        \State Repeat the above procedure for each sketch in $\Delta_{in}(u)$ and add entries into $I_{in}[T_{in,u}]$.
  \EndFor

  \State \textbf{return} $\widetilde{\Delta} = \{ I_{out}, I_{in} \}$ as the encrypted graph index.
  \end{algorithmic}
\end{algorithm}

\subsection{Privacy-preserving $\alpha$-CSD Querying}
In order to enhance protection of sensitive information,
we construct a privacy-preserving $\alpha$-CSD querying scheme $GraphEnc_{2}=(KeyGen, Setup, Query)$,
where the key generation procedure is the same as in $GraphEnc_{1}$, with improved index construction and CSD query procedures as exhibited in
Algorithms \ref{Algorithm:GraphEnc2:setup} and \ref{Algorithm:GraphEnc2:Query}, respectively.

\renewcommand{\algorithmicrequire}{\textbf{Input:}}
\renewcommand{\algorithmicensure}{\textbf{Output:}}

\begin{algorithm}[t]
  \footnotesize
  \caption{Setup algorithm for $GraphEnc_{2}$}\label{Algorithm:GraphEnc2:setup}
  \begin{algorithmic}[1]
  \Require A secret key $K$, a key pair $(pk,sk)$, an approximation ratio $\alpha$, an amplification factor $\phi$, and an original graph $G$.
  \Ensure The encrypted graph index $\widetilde{\Delta}$.

  \State Generate the 2HCLI $\Delta=\{\Delta_{out}, \Delta_{in}\}$ of $G$.
  \State Initialize two dictionary $I_{out}$ and $I_{in}$.
  \State Let $\mathcal{B}$ be the maximum distance over the sketches and set $N = 2 \mathcal{B} +1$.

  \For {each $u \in G$}
        \State Set $S_{out,u}=h(K, u||1)$, $T_{out,u}=h(K, u||2)$, $S_{in,u}=h(K, u||3)$, and $T_{in,u}=h(K, u||4)$.
        \State Initialize a counter $\omega=0$

        \For {each $(v, d_{u,v}, c_{u,v}) \in \Delta_{out}(u)$}
            \State Compute $V=h(K, v||0)$.
            \State Compute $D_{u,v} =$ SWHE.Enc$(pk, 2^{N-d_{u,v}})$.
            \State Compute $C_{u,v}=$ ORE.Enc$(K, \phi c_{u,v})$.

            \State Set $T_{out,u,v} = h(T_{out,u}, \omega)$ and $S_{out,u,v} = g(S_{out,u}, \omega)$.
            \State Compute $\Psi_{u,v}=S_{out,u,v} \oplus (V || D_{u,v} || C_{u,v})$.
            \State Set $I_{out}[T_{out,u,v}] = \Psi_{u,v}$.
            \State Set $\omega=\omega+1$.
        \EndFor

        \State Repeat the above procedure for each sketch in $\Delta_{in}(u)$
        and obtain $I_{in}[T_{in,u,v}]$,
        except that: (i) set $T_{in,u,v} = h(T_{in,u}, \omega)$ and $S_{in,u,v} = g(S_{in,u}, \omega)$, and
        (ii) compute $\Psi_{u,v}=S_{in,u,v} \oplus (V || D_{u,v} || C_{u,v})$.
  \EndFor

  \State \textbf{return} $\widetilde{\Delta} = \{ I_{out}, I_{in} \}$ as the encrypted graph index.
  \end{algorithmic}
\end{algorithm}

\renewcommand{\algorithmicrequire}{\textbf{Input:}}
\renewcommand{\algorithmicensure}{\textbf{Output:}}

\begin{algorithm}[htbp]
  \footnotesize
  \caption{Query algorithm for $GraphEnc_{2}$}\label{Algorithm:GraphEnc2:Query}
  \begin{algorithmic}[1]
  \Require The \emph{user}'s input are the secret key $K$, secret key pair $(pk, sk)$, an amplification factor $\Phi$,
  and the query $q=(s, t, \theta)$.
  The \emph{cloud server}'s input is the encrypted index $\widetilde{\Delta}$.
  \Ensure \emph{user}'s output is $dist_{q}$
  and \emph{cloud server}'s output is $\bot$.

  \State \emph{user} generates $S_{out,s}=h(K, s||1)$, $T_{out,s}=h(K, s||2)$,
  $S_{in,t}=h(K, t||3)$ and $T_{in,t}=h(K, t||4)$.
  \State \emph{user} constructs a cost constraint tree $T_{\theta}$ based on $\phi * \theta$ using secret $K$ as described in Section \ref{sec:Tree:Based:Ciphertexts:Comparison:Approach}.
  \State \emph{user} sends $\tau_{s,t} = (S_{out,s}, T_{out,s}, S_{in,t}, T_{in,t}, T_{\theta})$ to \emph{cloud server}.

  \State \emph{cloud server} parses $\tau_{s,t}$
  as $(S_{out,s}, T_{out,s}, S_{in,t}, T_{in,t}, T_{\theta})$.

  \State \emph{cloud server} initializes a set $L_{s}$ and a counter $\omega=0$.
  \State \emph{cloud server} computes $T_{out,s,v} = h(T_{out,s}, \omega)$.
  \While {$I_{out}[T_{out,s,v}] \ne \bot$}
        \State \emph{cloud server} computes $S_{out,s,v} = g(S_{out,s}, \omega)$.
        \State \emph{cloud server} performs $(V || D_{s,v} || C_{s,v}) = \Psi_{s,v} \oplus S_{out,s,v} $.
        \State \emph{cloud server} add $(V, D_{s,v}, C_{s,v})$ into $L_{s}$.
        \State Set $\omega=\omega+1$.
        \State \emph{cloud server} computes $T_{out,s,v} = h(T_{out,s}, \omega)$.
  \EndWhile

  \State \emph{cloud server} initializes a set $L_{t}$ and a counter $\omega=0$.
  \State \emph{cloud server} computes $T_{in,v,t} = h(T_{in,t}, \omega)$.
  \While {$I_{in}[T_{in,v,t}] \ne \bot$}
        \State \emph{cloud server} computes $S_{in,v,t} = g(S_{in,t}, \omega)$.
        \State \emph{cloud server} performs $(V || D_{v,t} || C_{v,t}) = \Psi_{v,t} \oplus S_{in,v,t} $.
        \State \emph{cloud server} add $(V, D_{v,t}, C_{v,t})$ into $L_{t}$.
        \State Set $\omega=\omega+1$.
        \State \emph{cloud server} computes $T_{in,v,t} = h(T_{in,t}, \omega)$.
  \EndWhile

  \State For each encrypted vertex identifier $v$ that appears in both in $L_{s}$ and $L_{t}$,
  the \emph{cloud server} performs the cost constraint filtering operation through Algorithm \ref{Algorithm:tree based ciphertexts comparison},
  and add the pair $(D_{s,v}, D_{v,t})$ which satisfies
  the cost constraint $\phi \theta$ into a set $Y$.
  The pair that Algorithm \ref{Algorithm:tree based ciphertexts comparison} cannot verify
  is also added into $Y$.

  \State For each pair in $Y$,
  the \emph{cloud server} first computes $d_{i} = $ SWHE.Eval$(\times, D_{s,v}^{i}, D_{v,t}^{i})$, and then computes $d = \sum_{i=1}^{|Y|}d_{i}$.

  \State \emph{cloud server} returns $d$ to the \emph{user}.
  \State \emph{user} decrypts $d$ with $sk$.

  \State \textbf{return} Decrypted value of $d$ as $dist_{q}$.
  \end{algorithmic}
\end{algorithm}

The \emph{Setup} for $GraphEnc_{2}$ works as follows.
The \emph{user} first builds the 2HCLI $\Delta$ of graph $G$, and then encrypts sketches associated with $u \in G$ (i.e., $\Delta_{out}(u)$ and $\Delta_{in}(u)$), as described in lines 2-17.

Note that in order to prevent the leakage of the sketch size in the previous straightforward approach,
we split each encrypted sketch $I_{out}(u)$ and $I_{in}(u)$, and ensure that they are stored in the dictionary separately, with a size of one.
More precisely, we utilize a counter $\omega$ and generate the unique $T_{out,u,v}$ and $S_{out,u,v}$ for each entity
in $\Delta_{out}(u)$ (line 11).
Similarly, the unique $T_{in,u,v}$ and $S_{in,u,v}$ for each entity in $\Delta_{in}(u)$ can be generated (line 16).
The $T_{out,u,v}$ (or $T_{in,u,v}$) indicates the position that this entity will be stored in $I_{out}$ (or $I_{in}$),
which ensures each position in the dictionary $I_{out}$ (or $I_{in}$) having only one entity.

$S_{out,u,v}$ (or $S_{in,u,v}$) is used to make an XOR operation with $(V || D_{u,v} || C_{u,v})$.
Since $S_{out,u,v}$ (or $S_{in,u,v}$) is different for each sketch, the XOR operation makes the resulting $\Psi_{u,v}$ indistinguishable,
which guarantees that the \emph{static} encrypted graph index $\widetilde{\Delta}$ reveals neither the number of common vertices between $I_{out}(u)$ and $I_{in}(v)$, nor the order information of costs.

The \emph{Query} in Algorithm \ref{Algorithm:GraphEnc2:Query} works as follows.
Assume that the \emph{user} asks for the shortest distance between $s$ and $t$,
whose total cost does not exceed $\theta$.
She first generates the query token $\tau_{s,t}$ and sends it to the \emph{cloud server} (lines 1-3).
Upon receiving the token $\tau_{s,t}$,
the \emph{cloud server} searches in the index and obtains $L_{s}$ and $L_{t}$ (lines 5-22).
That is, the \emph{cloud server} iteratively judges whether the dictionary $I_{out}$ ($I_{in}$)
contains the key $T_{out,s,v}$ ($T_{in,v,t}$) or not.
If it exists, then it adds the corresponding entity into the set $L_{s}$ ($L_{t}$).

Once $L_{s}$ and $L_{t}$ are obtained,
the \emph{cloud server} performs the cost constraint filtering (line 23) and computes $d$ (line 24),
which are the same as described in the straightforward approach.
Finally, the \emph{user} gets the final answer by decrypting $d$, which is returned by the \emph{cloud server}, using its $sk$.

\section{Tree-Based Ciphertexts Comparison Approach}\label{sec:Tree:Based:Ciphertexts:Comparison:Approach}
This section introduces a tree-based ciphertexts comparison approach,
which is used for cost constraint filtering in the graph encryption scheme described in Section \ref{sec:main scheme}.

\subsection{Scenarios}
Assume that there is a \emph{user} (i.e., $\mathcal{U}$) and a \emph{server} (i.e., $\mathcal{R}$).
$\mathcal{U}$ has many integers which are encrypted by a kind of cryptography algorithm
and then outsourced to $\mathcal{R}$.
Now, $\mathcal{U}$ wants to ask for $\mathcal{R}$ to obtain integer pairs, e.g., ($x$, $y$),
whose sum does not exceed $\theta$.
Note that the plaintexts of $x$, $y$ and $\theta$ could not be disclosed to $\mathcal{R}$, except for the greater-than, equality, or less-than relationship.
A naive approach is to download all the integers, calculate the summation locally,
and choose the integer pairs satisfying the constraint.
This method, however, is meaningless if one wants to offload the computation to the cloud.
Hence, it is desirable to have a practical solution to this problem.

Note that this scenario is different from
the well-known SMC scheme.
In the setting of SMC \cite{SMC_Ben2016, SMC_Ben2008FairplayMP}, 
a set of (two or more) parties with private inputs wish to compute a function of their inputs while revealing nothing but the result of the function, which is used for many practical applications, such as exchange markets.
SMC is a collaborative computing problem that solves the privacy preserving problem among a group of mutually untrusted participants.
The ciphertexts of all pairs of ($x$, $y$) and the cost constraint $\theta$ are outsourced to the cloud server,
which is responsible for the inequality tests.
Furthermore, we could reveal the relationship between the sum of two ciphertexts and another ciphertext to the \emph{server}, which is referred to as
\emph{controlled disclosure} in the literature \cite{Structured:encryption}.

It seems that we might leverage the homomorphic encryption technique, since it supports a sum operation of calculating $x + y$.
Nevertheless, as the homomorphic encryption is probabilistic,
we are unable to determine the relationship between $x + y$ and $\theta$ over their ciphertexts.

\subsection{Main Idea}
The main idea of the tree-based ciphertexts comparison protocol is to encode an integer with the ORE primitive.
To the best of our knowledge,
none of the existing approaches can support ORE and homomorphism properties simultaneously.
Hence, we design a novel method to address this problem,
which is motivated by the following facts.

If we want to compare $x+y$ with $\theta$,
we can compare $x$ with $\theta /2$ and $y$ with $\theta/2$, respectively.
Now, we result in 4 possible cases corresponding to combinations of the two relationships.
If $x > \theta /2$ ($x \le \theta /2$) and $y > \theta /2$ ($y \le \theta /2$),
we can know that $x+y > \theta$ ($x+y \le \theta $).
In the rest two cases, i.e., $x > \theta /2$ and $y < \theta /2$, or $x \le \theta /2$ and $y \ge \theta /2$,
we cannot achieve a deterministic result.
At this point,
we can further divide $\theta / 2$ into $\theta / 4$.
And then we can compare $x$ and $y$ with $\theta / 4$ and $3\theta / 4$, respectively.

By iteratively performing such an operation,
we can determine the relationship between $x+y$ and $\theta$ with an increasing probability.
Due to the ORE property,
it is easy to perform the above operations over ciphertexts.
Next, we will show how to implement this idea efficiently by utilizing a tree structure.

\subsection{Details of Protocol}
To implement the comparison of $x+y$ and $\theta$ over their ciphertexts,
we construct a \emph{cost constraint tree}, whose nodes represent specific values that are related to $\theta$.
For clarity, we define $E(m)$ as the ORE ciphertext of $m$.

\begin{figure}[t]
\begin{minipage}{0.49\textwidth}
  \centering
  \includegraphics[height=4cm]{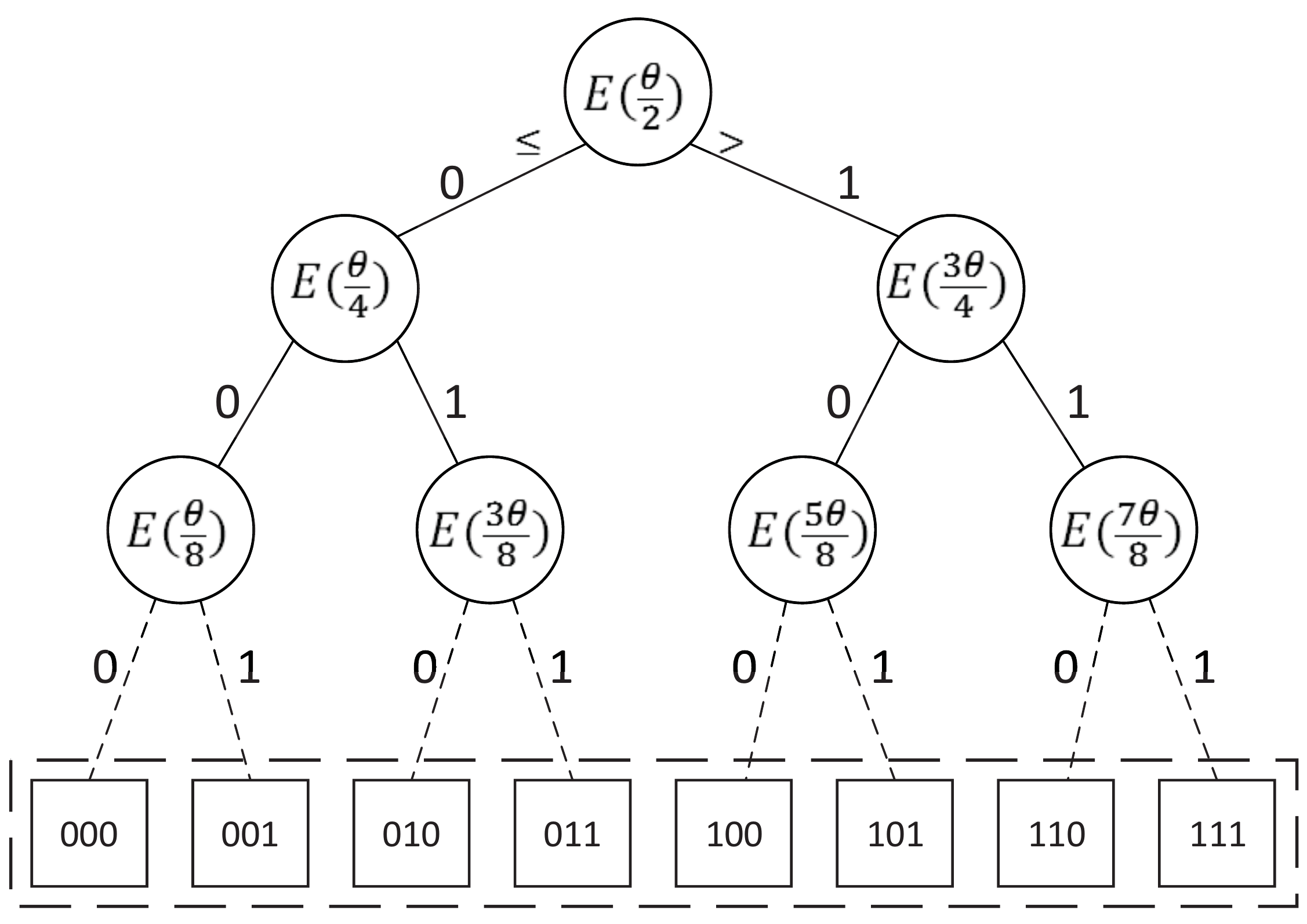}
\end{minipage}
\caption{An example of the cost constraint tree with a depth of 3, where circles represent nodes.
The boxes in the dashed rectangle indicate path codes for all possible comparison results. Note that these boxes are not a part of the tree. }\label{fig:tree_model}
\end{figure}

An example of the tree structure is depicted in Fig. \ref{fig:tree_model}.
For each node, we assign 0 to its left child path, while 1 to the right child path.
If an integer is not greater than the value of this node,
we take the left child path for further comparison; otherwise, we take the right child path.
Thus, for any path from the root node to a leaf node,
we can obtain a path code,
which is an effective representation of the comparison procedure.
For instance, an incoming integer $5\theta/16$ would traverse
Nodes $E(\theta/2)$, $E(\theta/4)$, and $E(3\theta/8)$, and thereby end with a path code of 010.
We define the length (i.e., the number of bits) of a path code as $\beta$.
Note that $\beta$ is actually equal to the depth of the tree which is denoted by $d_\theta$.

Now the relationship between $x + y$ and $\theta$ can be determined as follows.
We first get the ORE ciphertexts of $x$ and $y$,
as well as their path codes $c_{x}$ and $c_{y}$ by traversing the tree separately.
When computing $c_{x} + c_{y}$,
if an overflow occurs (i.e., $c_{x} + c_{y} \ge 2^{\beta}$),
we know that $x+y > \theta$ with confidence.
If $c_{x} + c_{y} \le 2^{\beta}-2$,
we also know that $x+y \le \theta$ with confidence.
Otherwise, we are unable to determine the relationship and end up with an \emph{uncertainty}.
We summarize this procedure in Algorithm \ref{Algorithm:tree based ciphertexts comparison}.

\renewcommand{\algorithmicrequire}{\textbf{Input:}}
\renewcommand{\algorithmicensure}{\textbf{Output:}}

\begin{algorithm}[htbp]
  \footnotesize
  \caption{Tree-Based Ciphertexts Comparison Algorithm}\label{Algorithm:tree based ciphertexts comparison}
  \begin{algorithmic}[1]
  \Require Two ORE ciphertexts $E(x)$, $E(y)$ and a cost constraint tree whose depth is $d_{\theta}$.
  \Ensure The relationship between $x+y$ and $\theta$.

  \State Initialize a counter $\omega=1$ and two empty strings $c_x$ and $c_y$.

  \While {$\omega \le d_{\theta}$}
        \State Visit the $\omega$-th level of the tree with $E(x)$ and concatenate $c_x$ with corresponding $0$ or $1$.
        \State Visit the $\omega$-th level of the tree with $E(y)$ and concatenate $c_y$ with corresponding $0$ or $1$.
        \State Set $\omega = \omega + 1.$
  \EndWhile

  \If {$c_{x} + c_{y} \ge 2^{\omega}$}
        \State \textbf{return} $>$.
  \EndIf
  \If {$c_{x} + c_{y} \le 2^{\omega} - 2$}
        \State \textbf{return} $\le$.
  \EndIf

  \State \textbf{return} {\emph{uncertainty}}.
  \end{algorithmic}
\end{algorithm}

\textbf{Discussion.}
Observe that when we go through a cost constraint tree,
one more step can further reduce the uncertainty of the relationship between $x+y$ and $\theta$ by half.
We denote the probability of \emph{uncertainty} as
  \begin{center}
    Pr$[\neg certainty]={(\frac{1}{2})}^{\beta}$.
  \end{center}
where $\beta$ is the length of the path code.
We can easily know the probability of certainty is
  \begin{center}
    Pr$[certainty] = 1-$ Pr$[\neg certainty]= 1 - {(\frac{1}{2})}^{\beta}$.
  \end{center}
When the tree depth is 6 (e.g., $\beta=6$),
the probability of certainty could reach about 0.9844.

Another observation is the comparison procedure
reveals the order information between $x$ (or $y$) and $\theta$. Thus, the \emph{server} can infer the interval that $x$ belongs to with precision of $2^{-\beta}$.
To prevent the \emph{server} from inferring the real value of
$x$, in \texttt{Connor},
the \emph{user} randomly picks a big integer number $\phi$ that is applied to $x$, $y$, and $\theta$ simultaneously,
which significantly enlarges the plaintext and ciphertext spaces (e.g., $2^{128}$).
The value of $\beta$ is generally a small integer (e.g., 6 in our implementation) that is determined by the \emph{user},
and both $\phi$ and $\theta$ are kept secret by the \emph{user}.
Therefore, the \emph{server} cannot infer the real value of $x$ (or $y$) from the order relationship among ciphertexts.
We will formally analyze the leakage functions and security issues in the next section.

\section{Complexity and Security Analyses}\label{sec:security}
This section presents the complexity and security analyses on the proposed graph encryption scheme \texttt{Connor}.

\subsection{Complexity Analysis}
\texttt{Connor} mainly consists of
the \emph{Setup} and \emph{Query} algorithms, as described in
Algorithms \ref{Algorithm:GraphEnc2:setup} and \ref{Algorithm:GraphEnc2:Query}.

The dominant component in determining the complexity of the \emph{Setup} algorithm is the encryption of the plain 2HCLI generated from a graph $G$.
Let $\mu$ be the total sketch for all vertices in $G$,
then the time complexity and space complexity are both $\mathcal{O}(n\mu)$, where $n$ is the number of vertices in $G$.


The \emph{Query} algorithm consists of a query token generation process on the \emph{user} side and a CSD query process  on the cloud \emph{server} side.
Let $\eta$ be the maximum size of the sketch associated with each vertex in $G$.
The complexity of the query token generation process is mainly determined by the construction of a cost constraint tree, whose time complexity and space complexity are both $\mathcal{O}(2^{d_{\theta}})$.
For the CSD querying process,
the time complexity of getting $L_{s}$ and $L_{t}$,
performing cost constraint filtering,
and performing distance computation are $\mathcal{O}(\eta)$,
$\mathcal{O}(\eta d_{\theta})$, and $\mathcal{O}({\eta})$, respectively.
The space complexity of the above three components are
$\mathcal{O}(\eta)$, $\mathcal{O}(\eta + 2^{d_{\theta}})$,
and $\mathcal{O}(\eta)$, respectively.
Therefore, the total time complexity and space complexity of
the CSD querying process are $\mathcal{O}({\eta} d_{\theta})$ and $\mathcal{O}(\eta + 2^{d_{\theta}})$, respectively.

\subsection{Security Analysis}

We now present the security analysis on \texttt{Connor}.
For clarity, we first discuss the leakage functions,
and then prove that \texttt{Connor} is secure under the CQA2-security model.

\textbf{Setup Leakage.}
The leakage function $\mathcal{L}_{Setup}$ of our construction reveals the information that can be deduced from the secure 2HCLI $\widetilde{\Delta}$ of graph $G$,
including the total number of vertices in the graph $n$,
the maximum distance over all the sketches
$\mathcal{B} = max_{u \in V} max_{\{(v, d_{u,v},c_{u,v}) \in \Delta_{out}, (v, d_{u,v},c_{u,v}) \in \Delta_{in}\}} d_{u,v}$, and the size of $\widetilde{\Delta}$.
More precisely, the size of $\widetilde{\Delta}$ consists of
the total number of sketch entities in $I_{out}$ and $I_{in}$, which are denoted by $\Omega_{out}$ and $\Omega_{in}$, respectively.
Thus, the leakage function $\mathcal{L}_{Setup} =
(n, \mathcal{B}, \Omega_{out}, \Omega_{in})$.

Note that the order relationship of pairwise costs and the order relationship between the cost and cost constraint are not included in $\mathcal{L}_{Setup}$,
because for each entity in sketches, we make an XOR operation using a unique integer value after we encrypt it, and this makes each entity in sketches are indistinguishable.

\textbf{Query Leakage.}
The leakage function $\mathcal{L}_{Query}$ of our construction
consists of the query pattern leakage, the sketch pattern leakage, and the cost pattern leakage.
Intuitively, the query pattern leakage
reveals whether a query has appeared before.
The sketch pattern leakage reveals the sketch associated to a queried vertex, the common vertices between two different sketches, and the size of the sketches of queried vertices.
The cost pattern leakage reveals 1) the order relationship among costs, and 2) the order relationship between costs and the cost constraint during the query procedure.
We formalize these leakage functions as follows.

\begin{myDef}(QUERY PATTERN LEAKAGE).
Let $\textbf{\emph{q}}=(q_{1}, q_{2}, \dots, q_{m})$ be a non-empty sequence of queries.
Each query $q_{i}$ specifies a tuple ($u_{i}$, $v_{i}$, $\theta_{i}$).
For any two queries $q_{i}$ and $q_{j}$,
define $Sim(q_{i}, q_{j})=(u_{i}=u_{j}, v_{i}=v_{j}, \theta_{i}=\theta_{j})$, i.e., whether each element of $q_{i}=(u_{i}, v_{i}, \theta_{i})$ matches each element of $q_{j}=(u_{j}, v_{j}, \theta_{j})$, respectively.
Then, the query pattern leakage function $\mathcal{L}_{QP}(\textbf{\emph{q}})$ returns an $m \times m$ (symmetric) matrix, in which each entry ($i$, $j$) equals $Sim(q_{i}, q_{j})$.
Note that $\mathcal{L}_{QP}(\textbf{\emph{q}})$ does not leak the identities of the query vertices.
\end{myDef}

\begin{myDef}(SKETCH PATTERN LEAKAGE).
Given a secure 2HCLI $\widetilde{\Delta}$ of a graph $G$ and
a query $q = (u, v, \theta)$, the sketch pattern leakage function $\mathcal{L}_{SP}(\widetilde{\Delta}, q)$ is defined as $(\Sigma, \Upsilon)$.
$\Sigma$ is a list, each element of which is the sketches associated to the queried vertices, and $\Upsilon$ is a pair $(X, Z)$, where $X={h(v):(v, d, c) \in I_{out}}$
and $Z={h(v):(v, d, c) \in I_{in}}$ are multi-sets and
$h: {\{0,1\}}^{\lambda} \times {\{0,1\}}^{*} \to {\{0,1\}}^{\lambda}$ is a particular pseudo-random function.
\end{myDef}

\begin{myDef}(COST PATTERN LEAKAGE).
The cost constraint $\theta$ in a query $q$ can essentially be represented by a certain number of uniform intervals.
Let $d_{\theta}$ be the depth of the cost constraint tree $T_{\theta}$ (c.f. Section \ref{sec:Tree:Based:Ciphertexts:Comparison:Approach}).
The intervals associated with $\theta$ are
$[{{(i-1)\theta} / 2^{d_{\theta}}}, {{i}\theta / 2^{d_{\theta}}}]$,
where $1 \le i \le 2^{d_{\theta}}$.
Assign each interval with a list $\mu$,
i.e., the $i$-th interval is associated with $\mu_{i}$,
which stores all the cost values belong to this interval.
The leaked interval information forms an array $Arr$,
of which the $i$-th element is $\mu_{i}$ (i.e., $Arr[i] = \mu_{i}$).
In addition, assume that $z$ is the total number of entries in
the sketches of the queried vertices.
For each pair of costs $c_{i}$ and $c_{j}$, its order relationship of the greater-than, equality, and less-than can be represented by $1$, $0$, and $-1$, respectively.
The leaked order information of costs is a $z \times z$ (symmetric) matrix $\nabla$ with each entry ($i$, $j$) being  $1$, $0$, or $-1$.
Therefore, the cost pattern leakage function
$\mathcal{L}_{CP}(\widetilde{\Delta}, q)=(Arr, \nabla)$.
\end{myDef}

Thus, $\mathcal{L}_{Query}=(\mathcal{L}_{QP}(\textbf{\emph{q}}), \mathcal{L}_{SP}(\widetilde{\Delta}, q), \mathcal{L}_{CP}(\widetilde{\Delta}, q))$.

The leakage functions are defined over the 2HCLI rather than the original graph.
In fact, the information leakage of the original graph is limited to the minimum number of paths for the queried source-destination vertices.
It can be defined as an $n \times n$ (symmetric) matrix $\Lambda$, where $n$ is the number of vertices in the graph.
Each element in $\Lambda$ is NULL, 0, or a positive integer,
which indicates an uncertain status (i.e., topology is well protected), disconnection, or the minimum number of paths
of the two queried vertices, respectively.

For the cost values in the 2HCLI,
we introduce a \emph{user}-held amplification factor $\phi$ to enlarge the plaintext and ciphertext spaces.
Thus, the \emph{server} cannot infer the real cost values just from their order information revealed by the leakage function $\mathcal{L}_{CP}(\widetilde{\Delta}, q))$.
For the distance values in the 2HCLI,
we use the SWHE encryption to protect their real values from the \emph{server}.


\textbf{Theorem 1.} \emph{If the cryptography primitives $g$, $h$, ORE, and the SWHE are secure,
then the proposed graph encryption scheme $\Pi =(KeyGen, Setup, Query)$
is $(\mathcal{L}_{Setup}, \mathcal{L}_{Query})$-secure against the adaptive chosen-query attack.}

\begin{proof}
The key idea is constructing a simulator $\mathcal{S}$.
Given the leakage functions $\mathcal{L}_{Setup}$ and $\mathcal{L}_{Query}$,
$\mathcal{S}$ constructs a fake encrypted 2HCLI structure
$\widetilde{\Delta}^{*} = \{ I_{out}^{*}, I_{in}^{*} \}$
and a list of query $q^{*}$.
If for all PPT adversaries $\mathcal{A}$,
they cannot distinguish between the two games \textbf{Real} and \textbf{Ideal},
we can say that our graph encryption scheme is $(\mathcal{L}_{Setup}, \mathcal{L}_{Query})$-secure against the adaptive chosen-query attack.

\textbf{Simulating} $\widetilde{\Delta}^{*}$.
  $\mathcal{S}$ handles each vertex $u_{i}$ ($1 \le i \le n$) to generate a fake $I_{out}^{*}$ in 2HCLI based on the leakage function $\mathcal{L}_{Setup}$.
  $\mathcal{S}$ randomly chooses $w_{i}$ for $u_{i}$ with $\sum_{1}^{n}w_{i} =\Omega_{out}$,
  and samples $l_{i} \gets \{{0,1\}}^{\lambda}$
  and $\eta_{i} \gets \{{0,1\}}^{\lambda}$ uniformly without repetition.
  For all $0 \le i < w_{i}$,
  $\mathcal{S}$ takes the following steps to simulate each sketch:
  $\mathcal{S}$ computes $l_{w_{i}}=h(l_{i}, w_{i})$ and
  $\eta_{w_{i}}=h(\eta_{i}, w_{i})$,
   where $h$ is a particular pseudo-random function.
  Then, it encrypts each vertex $v$ in the sketch of $u_{i}$ by computing $V^{*}=h(K^{*},v||0)$, where $K^{*}$ is a fake secret key.
  It randomly generates two integers $d$ and $c$
  and obtains ciphertexts $D^{*}$ and $C^{*}$ by encrypting $2^{N-d}$ ($N=2\mathcal{B}+1$) and $c$ using the SWHE and ORE schemes.
  Let $\Psi_{i}^{*} = \eta_{w_{i}} \oplus (V^{*}||D^{*}||C^{*})$.
  $\mathcal{S}$ stores $\Psi^{*}$ in the index $I_{out}^{*}$.
  That is, $I_{out}^{*}[l_{w_{i}}]= \Psi_{i}^{*}$.
  Similarly, $\mathcal{S}$ generates a fake $I_{in}^{*}$
  and finally obtains the fake 2HCLI $\widetilde{\Delta}^{*}=\{I_{out}^{*}, I_{in}^{*}\}$.

Simulating $q^{*}$.
  Given the leakage function $\mathcal{L}_{Query}=(\mathcal{L}_{QP}(q), \mathcal{L}_{SP}(\widetilde{\Delta}, q), \mathcal{L}_{CP}(\widetilde{\Delta}, q))$,
  $\mathcal{S}$ simulates the query token as follows.
  $\mathcal{S}$ first checks if either of the queried vertices $s$ and $t$ has appeared in any previous query.
  If $s$ appeared previously,
  $\mathcal{S}$ sets $S_{out,s}^{*}$ and $T_{out,s}^{*}$ to the values that were previously used.
  Otherwise,
  it sets $T_{out,s}^{*}=l_{i}$ and $S_{out,s}^{*}=\eta_{i}$ for some previously unused $l_{i}$ and $\eta_{i}$.
  It then remembers the association among $\eta_{i}$, $l_{i}$, and $s$.
  $\mathcal{S}$ takes the same steps for the queried vertex $t$: setting $S_{in,t}^{*}$ and $T_{in, t}^{*}$ analogously and associating $t$ with the selected $\eta_{i}$ and $l_{i}$.

  To simulate a fake cost constraint tree $T_{\theta}^{*}$,
  $\mathcal{S}$ first checks if the queried $\theta$ appeared in any previous query.
  If $\theta$ appeared previously,
  $\mathcal{S}$ sets the $T_{\theta}^{*}$ to the value that was previously used.
  Otherwise, $\mathcal{S}$ constructs a full binary tree based $\theta$ and encrypts each tree node by using the ORE scheme with a randomly generated key.
  $\mathcal{S}$ returns this encrypted tree as $T_{\theta}^{*}$.

  $\mathcal{S}$ simulates the query procedure as follows.
  Given the query token $(S_{out,s}^{*}, T_{out,s}^{*}, S_{in,t}^{*}, T_{in,t}^{*}, T_{\theta}^{*})$,
  $\mathcal{S}$ first checks if the query has been queried before.
  If yes, $\mathcal{S}$ returns the value that was previously used as the query result.
  Otherwise, $\mathcal{S}$ checks whether the queried vertex $s$ (or $t$) has been queried before.
  If the query vertex $s$ has appeared in a previous query,
  $\mathcal{S}$ sets $L_{s}^{*}$ to the values that were previously used from $\Sigma$ of $\mathcal{L}_{SP}(\widetilde{f}, q)$.
  Otherwise, for a newly appeared vertex $s$,
  $\mathcal{S}$ takes the following steps:
  To generate the sketches associated with $s$,
  $\mathcal{S}$ first initializes a set $L_{s}^{*}$ and a counter $\omega^{*}=0$,
  Then, it iteratively computes $T_{out,s,v}^{*}=h(T_{out,s}^{*}, w_{*})$ and $S_{out,s,v}^{*}=g(S_{out,s}^{*}, w_{*})$, and adds the tuple $(V^{*}, D_{s,v}^{*}, C_{s,v}^{*})$ into $L_{s}^{*}$,
  until $I_{out}^{*}[T_{out,s,v}^{*}]$ does not exist,
  where $(V^{*}, D_{s,v}^{*}, C_{s,v}^{*}) = I_{out}^{*}[T_{out,s,v}^{*}] \oplus S_{out,s,v}^{*}$.
  Similarly, $\mathcal{S}$ obtains the set $L_{t}^{*}$ for vertex $t$.
  Upon obtaining $L_{s}^{*}$ and $L_{t}^{*}$,
  $\mathcal{S}$ performs cost constraint filtering operation based on $T_{\theta}^{*}$ to get the candidate set $Y^{*}$.
  The theorem then follows from the CPA-security of SWHE.
  That is, $\mathcal{S}$ performs the SWHE computation over $Y^{*}$ and returns the query result.

Since the cryptography primitives $g$, $h$, ORE, and SWHE are secure,
the fake 2HCLI structure $\widetilde{\Delta}^{*}$ and the query sequence $q^{*}$
are indistinguishable from the real ones.
Therefore, for all PPT adversaries $\mathcal{A}$,
they cannot distinguish between the two games \textbf{Real} and \textbf{Ideal}.
Thus, we have
\begin{eqnarray*}
|\textbf{Pr}[Real_{\Pi, \mathcal{A}}(\lambda) = 1] - \textbf{Pr}[Ideal_{\Pi, \mathcal{A}, \mathcal{S}}(\lambda) = 1]| \le negl(\lambda).
\end{eqnarray*}
 where $negl(\lambda)$ is a negligible function.
\end{proof}

\section{Performance Evaluation}\label{sec:evaluation}
This section presents the evaluation of our graph encryption scheme through experiments on real-world datasets.

\subsection{Setup}
\textbf{Testbed.}
We implement the method introduced in \cite{Effective:Indexing:Approximate:Constrained:Shortest:Path:Queries}
for building the 2HCLI.
The ORE and SWHE in our implementation follow the methods described in \cite{Practical:Order:Revealing:Encryption:with:Limited:Leakage}
and \cite{boneh2005evaluating}, respectively.
The GMP library is used for big integer arithmetic.
We set the security parameter $\lambda=128$ and use the OpenSSL library for all the basic cryptographic primitives.
All the algorithms in our experiment are implemented in C++.
The experiments are conducted on a desktop PC equipped with Intel Xeon processor at 2.6 GHz and 8 GB RAM.

\textbf{Graph sets.}
The datesets used in our experiments are listed in Table \ref{tab:datasets}.
All these datasets are publicly available from the Standford SNAP website$\footnote{http://snap.stanford.edu/data/}$
and modeled as directed graphs.
For the datasets soc-Epinions1 and Email-EuAll,
we randomly select their subsets to make the index construction feasible with the limited computational resources.
Since these graphs are unweighted,
we generate a distance and a cost for each edge,
the value of which follows a uniform distribution between 1 and 100.
The cost criterion is used as the constraint.

\begin{table}[t]
\renewcommand{\arraystretch}{1.1}
\caption{The graph datasets used in our experiments} \label{tab:datasets}
\centering
\footnotesize{
\begin{tabular}{||C{2.0cm}|C{1.2cm}|C{1.2cm}|C{1.2cm}||}
\hline
Dataset  & Nodes  & Edges  & Storage \\ \hline
\hline
Email-EuAll & 21,721 & 34,351 &  335KB \\ \hline
soc-Epinions1 & 6,506 & 47,062 & 418KB \\ \hline
p2p-Gnutella25 & 22,687 & 54,705 & 632KB \\ \hline
p2p-Gnutella04 & 10,876 & 39,994 & 422KB \\ \hline
\end{tabular}}
\end{table}

\textbf{Methods to compare.}
Since this is the first work to address the CSD querying problem over encrypted graphs,
we compare our method with the one over unencrypted graphs.
We implement such a method following
the state-of-the-art method over plaintext graphs introduced in \cite{Effective:Indexing:Approximate:Constrained:Shortest:Path:Queries}.
The only difference is that we construct 2HCLI over the original graph, instead of an overlay graph.
As a result, our implementation has a higher query efficiency
but leads to a higher complexity of the index construction.

\textbf{Query sets.}
We randomly generate 200 queries over each dataset.
The origin $s$ and destination $t$ in each query are also randomly selected.
The cost constraint $\theta$ for each $(s,t)$ pair is set as follows.
We denote the \emph{lower} bound $c_{min}$ as the minimum cost of all paths from $s$ to $t$,
and the \emph{upper} bound $c_{max}$ as the minimum cost of the paths with the shortest distance from $s$ to $t$.
If the cost constraint $\theta < c_{min}$,
there will be no feasible answer to the query;
and if the cost constraint $\theta > c_{max}$,
the shortest distance is always a valid answer to the query.
To mitigate the impact of $\theta$ on the performance,
we randomly choose 50 values of $\theta$ for each query,
which falls in the interval $[c_{min}, c_{max}]$.

Another important parameter is $\alpha$, which determines the approximation guarantees of $\alpha$-CSD queries.
Since $\alpha$ is a constant value for all queries, we view it as a system parameter rather than part of specific queries.
In order to achieve a balance between query accuracy and system efficiency,
we set the approximation ratio $\alpha = 1.5$ for all queries.

\subsection{Evaluation of Secure 2HCLI and Query Token}
\textbf{Index Size and Construction Time.}
The index construction of the graph is a one-time and offline computation.
This process consists of two steps:
one is constructing the plain 2HCLI,
which is the same as the index construction process of the original plain CSD querying,
and the other is encrypting the plain 2HCLI,
which is the focus of this paper.
Therefore,
we consider the outputs of the first step as the index of unencrypted graph.

The index size and construction time are depicted in Table \ref{tab:index}.
Note that the index size and construction time of different datasets have a great difference, which is mainly caused by the difference in graph topologies.
Different from the original shortest distance query,
where there is only one shortest path between any two vertices,
in the CSD querying problem,
there usually exist multiple constrained shortest paths between any two vertices.
Intuitively,
a dense graph may bear a higher index construction cost than a sparse one.

In general, the size of each encrypted index is roughly 6$\times$ larger than that of the corresponding plain index.
The most important observation is that the index construction time of encrypted graphs
is slightly higher than the one of unencrypted graphs.
Thus, the key point of improving the index construction efficiency over an encrypted graph
is accelerating the process of constructing the plain 2HCLI of that graph.
We leave this attempt as the future work.

\begin{table}[t]
\small
\renewcommand{\arraystretch}{1.1}
\caption{Summary of Index Construction Cost} \label{tab:index} \centering
\centering
\footnotesize{
\begin{tabular}{|C{2.3cm}|C{1.0cm}|C{1.0cm}|C{1.0cm}|C{1.0cm}|}
\hline
\multirow{2}{*}{Metrics} &
\multicolumn{2}{c|}{Plain Graph Query} &
\multicolumn{2}{c|}{\texttt{Connor}} \\
\cline{2-5}
 &Time (mins)& Size (MB) & Time (mins) & Size (MB) \\ \hline
D1: Email-EuAll & 862.03 & 8.99 & 869.84 & 48.14 \\ \hline
D2: soc-Epinions1 & 7093.25 & 5.76 & 7098.44 & 32.79 \\ \hline
D3: p2p-Gnutella25 & 4206.96 & 138.50 & 4306.31 & 514.46 \\ \hline
D4: p2p-Gnutella04 & 3007.91 & 63.12 & 3054.55 & 297.95 \\ \hline
\end{tabular}}
\end{table}

\textbf{Query Token Generation.}
The construction of query tokens is independent of specific graphs,
we now analyze the size and generation time of a query token.
The query token mainly consists of 5 elements, namely
$S_{out,s}$, $T_{out,s}$, $S_{in, t}$, $T_{in,t}$, and $T_{\theta}$.
Each of the first 4 elements has a length of 16 bytes.
Since the size of each ORE ciphertext is 16 bytes,
a cost tree $T_{\theta}$ whose depth is $d_{\theta}$ has a size of $16 \times (2^{d_{\theta}}-1)$  bytes.
Therefore, the total size of a query token is $16 \times (2^{d_{\theta}} + 3)$ bytes.
Since $d_{\theta}$ is a relatively small value,
the size of a query token is usually less than $1$ KB.
The query token generation time with varying $d_{\theta}$
is depicted in Table \ref{tab:TokenGenTime}.
Although the query token generation time increases significantly with $d_{\theta}$,
the time cost is moderate for general cases (e.g., when $d_{\theta} \leq 6$).

\begin{table}[h]
\small
\renewcommand{\arraystretch}{1.1}
\caption{The query token generation time for different $d_{\theta}$} \label{tab:TokenGenTime} \centering
\centering
\footnotesize{
\begin{tabular}{|C{1.2cm}|C{0.4cm}|C{0.4cm}|C{0.4cm}|C{0.4cm}|C{0.4cm}|C{0.4cm}|C{0.41cm}|C{0.42cm}|}
\hline
$d_{\theta}$ & 1 & 2 & 3 & 4 & 5 & 6 & 7 & 8 \\ \hline
Time (ms) & 0.15 & 0.31 & 0.65 & 1.22 & 2.46 & 4.95 & 9.88 & 19.77 \\ \hline
\end{tabular}}
\end{table}

\subsection{Evaluation of Query Efficiency and Accuracy}
\textbf{Query Efficiency.}
To evaluate the query efficiency,
for each $\theta$,
we generate the cost constraint tree with a different depth $d_{\theta}$.
The query time is defined as the time interval from the submission of a query token to the receival of its query results.
We compute the average query time of 200 queries.

\begin{figure}[h]
\begin{minipage}{0.49\textwidth}
  \centering
  \includegraphics[height=4.5cm]{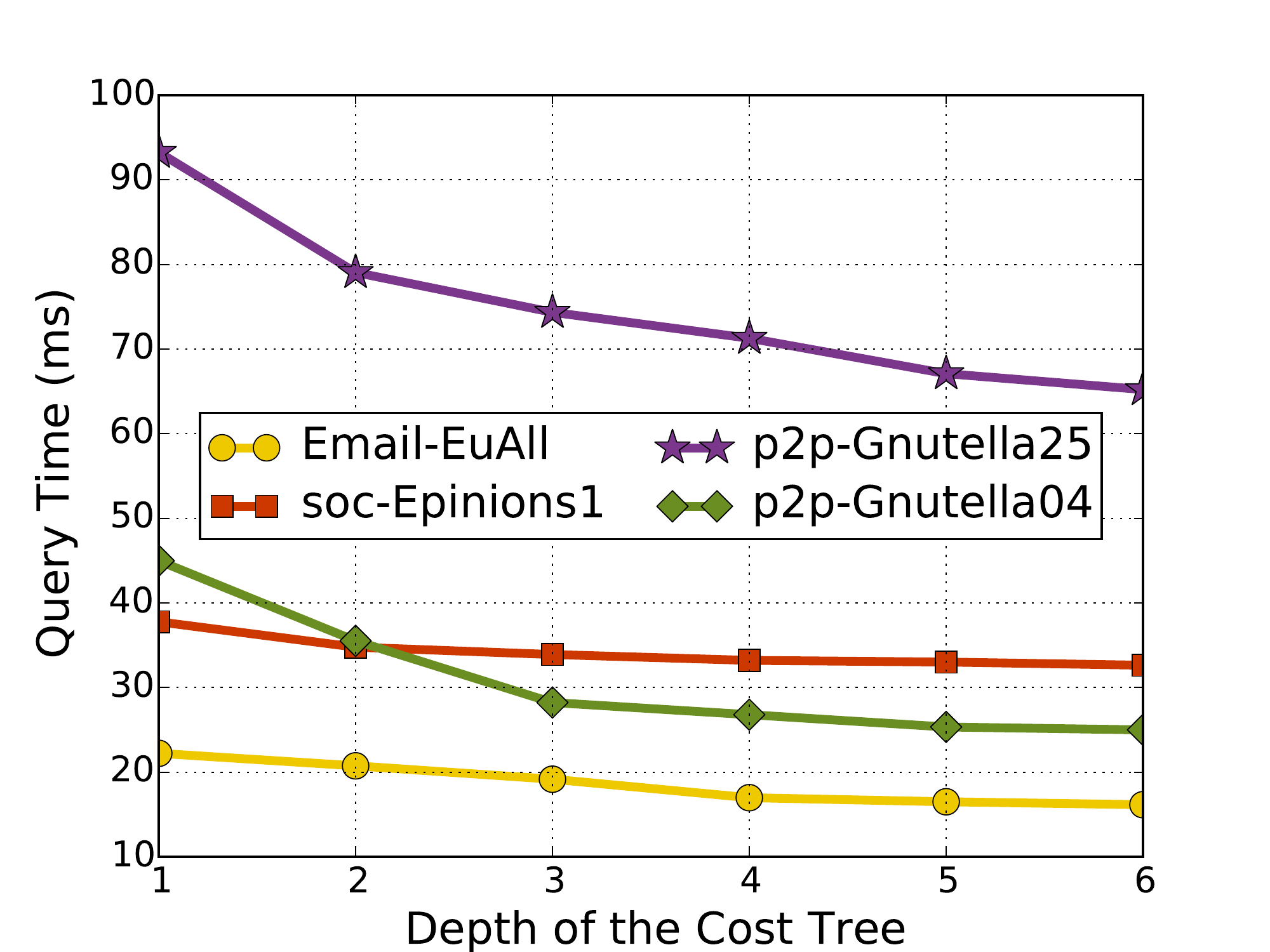}
\end{minipage}
\caption{The query time over encrypted 2HCLI with varying $d_{\theta}$.}\label{fig:QueryTimeForDepth}
\end{figure}

The average query time with varying $d_{\theta}$ over the encrypted 2HCLI is depicted in Fig. \ref{fig:QueryTimeForDepth}, where $d_{\theta}$ increases from 1 to 6.
We can see that the query time varies a lot for different graph datasets.
For each dataset, increasing $d_{\theta}$ can result in a decrease in the query time.
This is because a larger $d_{\theta}$ can filter out more distance pairs exceeding the cost constraint and thereby reduce the number of candidates for distance computation using SWHE, which is the dominant operation in time consumption.

\begin{figure*}[t]
\begin{minipage}{0.32\textwidth}
  \centering
  \includegraphics[height=4.5cm]{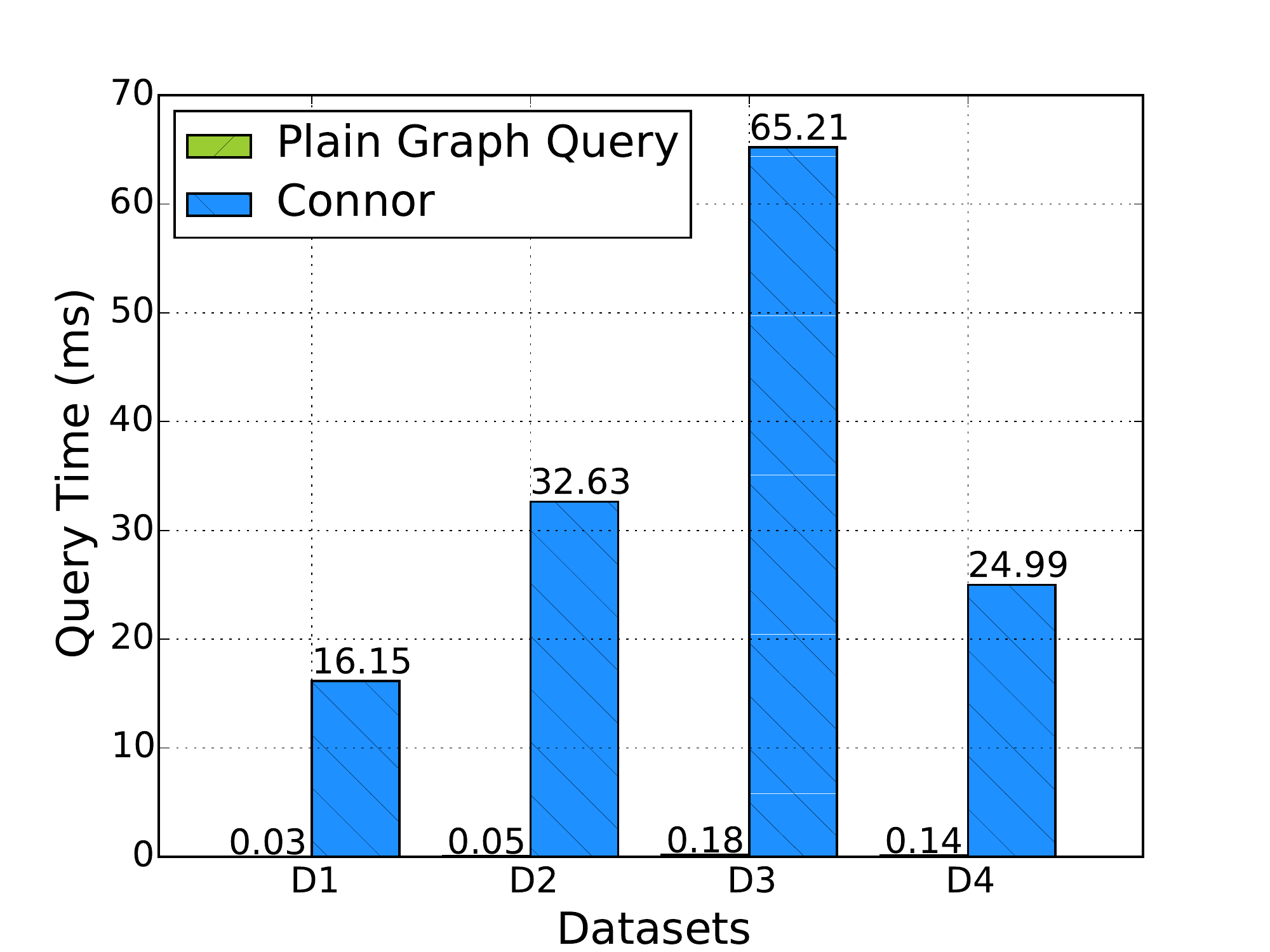}
  \caption{The query time over the plain 2HCLI and the encrypted 2HCLI ($d_{\theta} = 6$).}\label{fig:QueryTimeForCmp}
\end{minipage}
\quad
\begin{minipage}{0.32\textwidth}
  \centering
  \includegraphics[height=4.5cm]{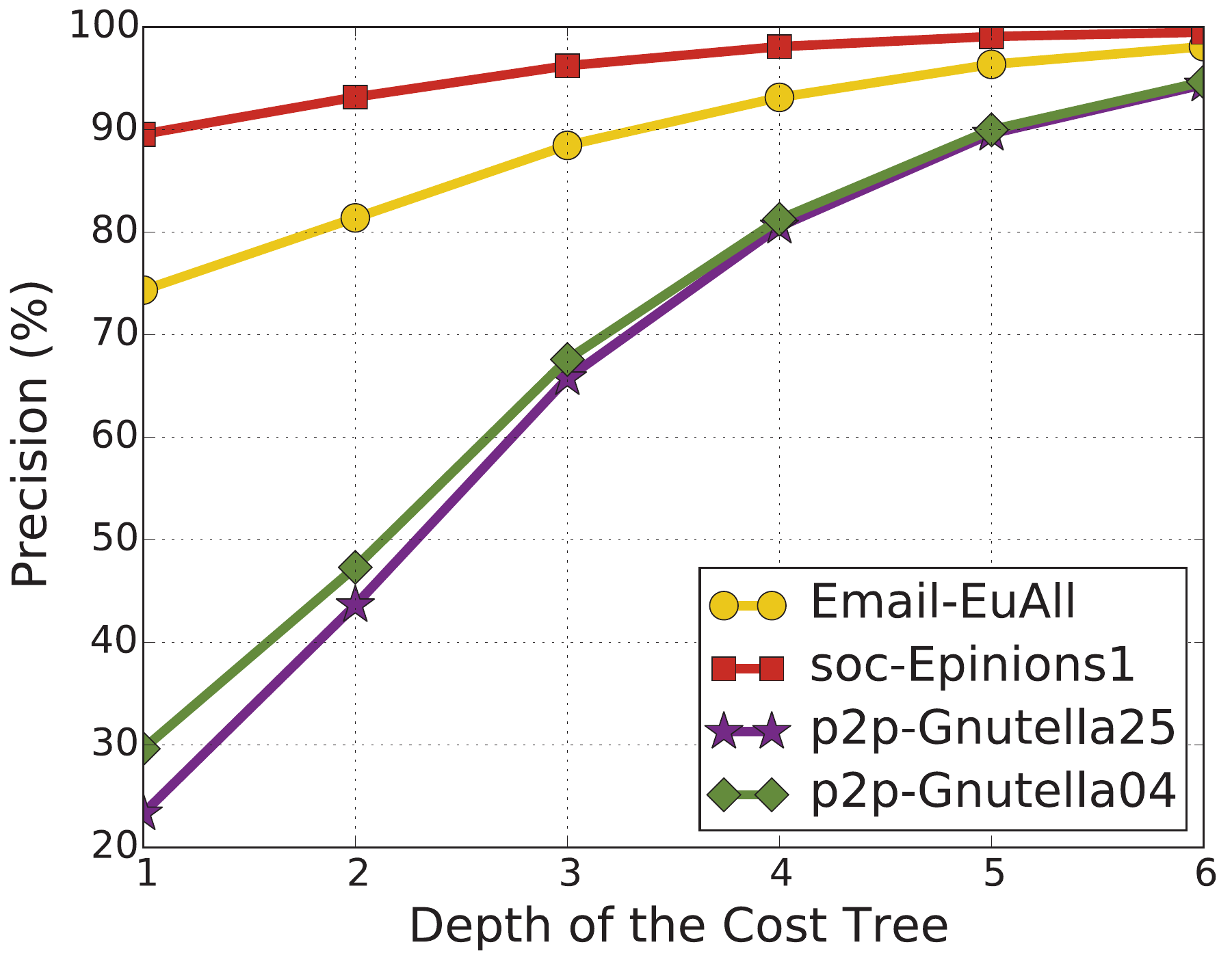}
    \caption{The query precision for different depth $d_{\theta}$ of the cost tree.}\label{fig:QueryAccuracyCost}
\end{minipage}
\quad
\begin{minipage}{0.32\textwidth}
  \centering
  \includegraphics[height=4.5cm]{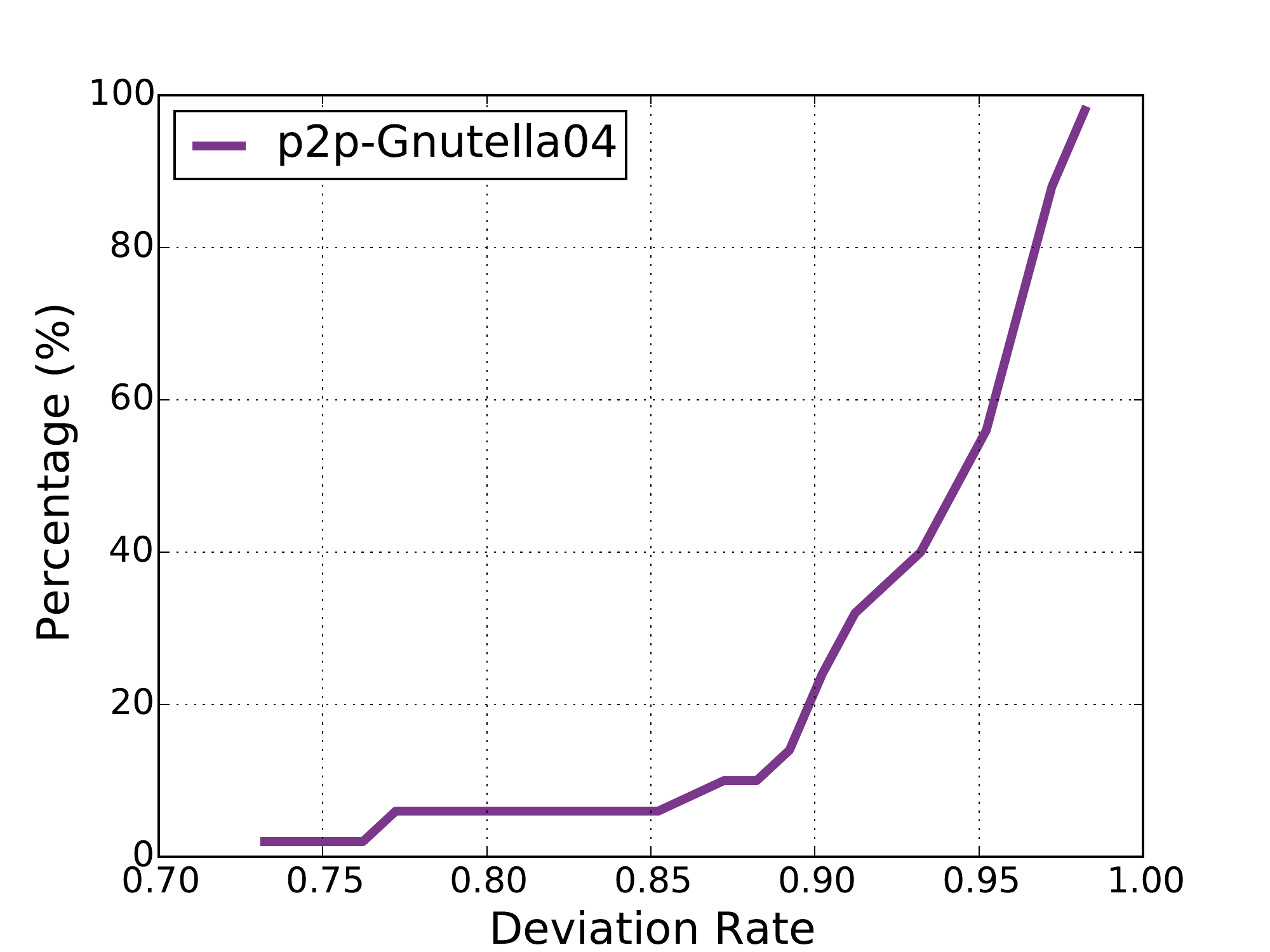}
    \caption{The CDF of deviation rate for different query ($d_{\theta} = 6$).}\label{fig:QueryAccuracyR}
\end{minipage}
\end{figure*}

Fig. \ref{fig:QueryTimeForCmp} presents the query time in the plain and encrypted scenarios for different datasets.
The query time over the encrypted 2HCLI is higher than that over the plain 2HCLI because of the time-consuming operations on ciphertexts (e.g., the cost filtering and distance computation).
Also, the time complexity of these operations is closely related to the size of a graph index listed in Table \ref{tab:index},
which leads to the difference among four datasets in
Fig. \ref{fig:QueryTimeForCmp}.


\textbf{Query Accuracy}.
In \texttt{Connor},
there are two components that affect the query accuracy,
namely the tree-based ciphertexts comparison
and the distance computation.
The former may keep some distance pairs that do not satisfy the cost constraint in the candidate set $Y$,
while the latter leverages the property of SWHE to obtain an approximate, but not exact, shortest distance based on all candidates in $Y$.

%

We use the well-known metric \emph{Precision} ($\mathcal{P}$)
to evaluate the accuracy of the cost constraint filtering process.
$\mathcal{P} = \frac{T_{p}}{T_{p}+F_{p}}$,
where $T_{p}$ and $F_{p}$ represent the numbers of distance pairs in $Y$ whose costs truly satisfy or exceed the cost constraint, respectively.
We use the same query as introduced above, and compute the $\mathcal{P}$ for each query. Finally, we can obtain the average precision $\bar{\mathcal{P}}$ for all the queries.

Fig. \ref{fig:QueryAccuracyCost} presents the relationship between the query precision $\bar{\mathcal{P}}$ and the depth of the cost constraint tree $d_{\theta}$ over different datasets.
We can see that for all the datasets,
$\bar{\mathcal{P}}$ increases with $d_{\theta}$,
because the cost constraint tree with a larger depth $d_{\theta}$
helps us to detect constraint violations with a higher probability,
as discussed in Section \ref{sec:Tree:Based:Ciphertexts:Comparison:Approach}.
In particular,
$\bar{\mathcal{P}}$ is more than $94\%$ for all datasets when $d_{\theta} = 6$.

To evaluate the accuracy of the final query results,
we propose a metric named the \emph{deviation rate}.
Let $r_{e}$ and $r_{p}$ be the query results returned by \texttt{Connor} and the algorithm over the corresponding plain graphs, respectively.
Then, we define the \emph{deviation rate} $\xi = r_{e} / r_{p}$, which indicates how far $r_{e}$ deviates from $r_{p}$.
Obviously,
a \emph{deviation rate} closer to 1 depicts more accurate query results.

Fig. \ref{fig:QueryAccuracyR} presents the cumulative distribution functions (CDFs) of the \emph{deviation rate} over the dataset p2p-Gnutella04.
We can see that $\xi$ is larger than $0.90$ for over $80\%$ of the query results, and larger than $0.73$ in the worst cases.
Therefore, \texttt{Connor} is capable of achieving a relatively high accuracy with moderate computation complexity.


\section{Conclusion}\label{sec:conclusion}
In this paper, we have presented \texttt{Connor}, the first graph encryption scheme that enables the cloud-based approximate CSD queries.
In particular, we proposed a tree-based ciphertexts comparison protocol for cost constraint filtering with controlled disclosure.
The security analysis showed that \texttt{Connor} could achieve the CQA2-security.
We implemented a prototype and evaluated the performance using the real-world graph datasets.
The evaluation results demonstrated the effectiveness of \texttt{Connor}.
In the future work, we plan to design techniques to support dynamic index updates.






%

\bibliographystyle{unsrt}
\bibliography{main}

%


\begin{IEEEbiography}
[{\includegraphics[width=0.8in,height=1in,clip,keepaspectratio]{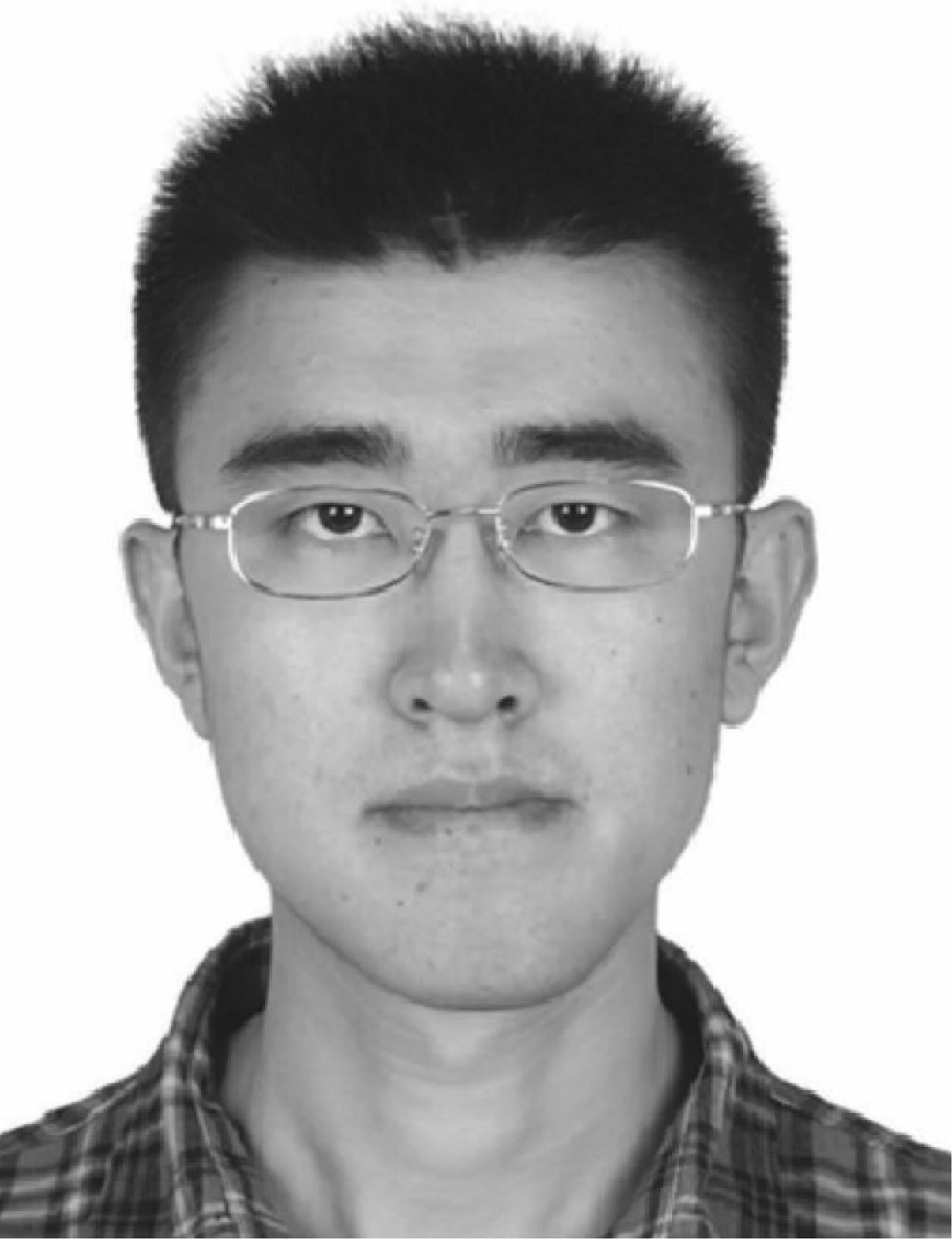}}]
{Meng Shen} received the B.Eng degree from Shandong University, Jinan, China in 2009, and the Ph.D degree from Tsinghua University, Beijing, China in 2014, both in computer science. Currently he serves in Beijing Institute of Technology, Beijing, China, as an assistant professor. His research interests include privacy protection of cloud-based services, network virtualization and traffic engineering.
He received the Best Paper Runner-Up Award at IEEE IPCCC 2014.
He is a member of the IEEE.
\end{IEEEbiography}

\vspace{-40pt}

\begin{IEEEbiography}
[{\includegraphics[width=0.8in,clip,keepaspectratio]{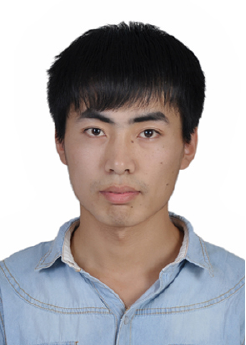}}]
{Baoli Ma} received the B.Eng degree in computer science from Beijing Institute of Technology, Beijing, China in 2015.
Currently he is a master student in the School of Computer Science, Beijing Institute of Technology.
His research interest is secure searchable encryption.
\end{IEEEbiography}

\vspace{-40pt}
%
\begin{IEEEbiography}
[{\includegraphics[width=0.8in,clip,keepaspectratio]{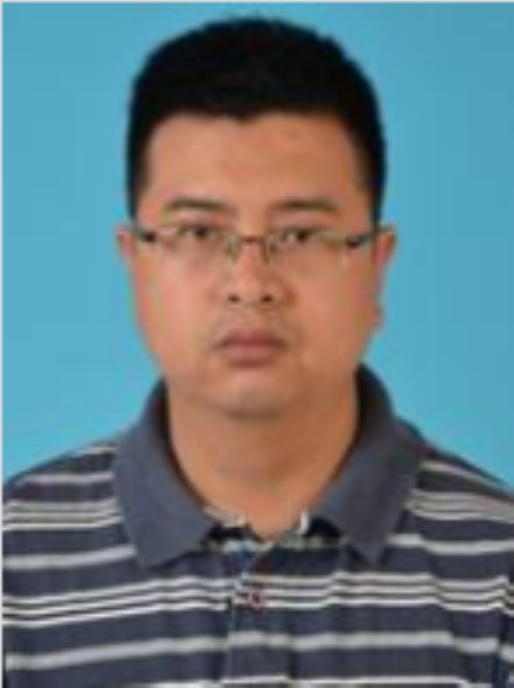}}]
{Liehuang Zhu} is a professor in the School of Computer Science, Beijing Institute of Technology.
He is selected into the Program for New Century Excellent Talents in University from Ministry of Education, P.R. China.
His research interests include Internet of Things, Cloud Computing Security, Internet and Mobile Security.
\end{IEEEbiography}

\vspace{-40pt}

\begin{IEEEbiography}
[{\includegraphics[width=0.8in,clip,keepaspectratio]{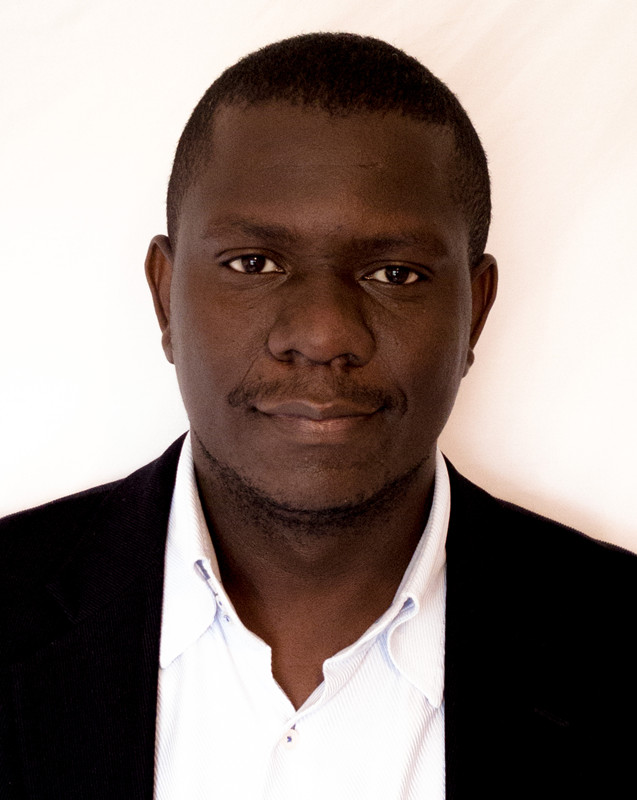}}]
{Rashid Mijumbi} received a PhD in telecommunications engineering from the Universitat Politecnica de Catalunya (UPC), Barcelona, Spain. He was a Post-Doctoral Researcher with the UPC and with the Telecommunications Software and Systems Group, Waterford, Ireland, where he participated in several Spanish national, European, and Irish National Research Projects. He is currently a Software Systems Reliability Engineer with Bell Labs CTO, Nokia, Dublin, Ireland. His current research focus is on various aspects of 5G, NFV and SDN systems. He received the 2016 IEEE Transactions Outstanding Reviewer Award recognizing outstanding contributions to the IEEE Transactions on Network and Service Management. He is a Member of IEEE.
\end{IEEEbiography}

\vspace{-80pt}

\begin{IEEEbiography}
[{\includegraphics[width=0.8in,clip,keepaspectratio]{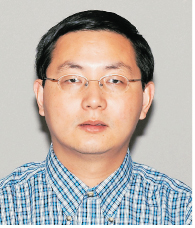}}]
{Xiaojiang Du} is a tenured professor in the Department of Computer and Information Sciences at Temple University, Philadelphia, USA.
Dr. Du received his B.S. and M.S. degree in electrical engineering from Tsinghua University, Beijing, China in 1996 and 1998, respectively. He received his M.S. and Ph.D. degree in electrical engineering from the University of Maryland College Park in 2002 and 2003, respectively.
His research interests are wireless communications, wireless networks, security, and systems.
He has authored over 200 journal and conference papers in these areas, as well as a book published by Springer.
Dr. Du has been awarded more than \$5 million US dollars research grants from the US National Science Foundation (NSF), Army Research Office, Air Force, NASA, the State of Pennsylvania, and Amazon.
He won the best paper award at IEEE GLOBECOM 2014 and the best poster runner-up award at the ACM MobiHoc 2014.
He serves on the editorial boards of three international journals.
Dr. Du is a Senior Member of IEEE and a Life Member of ACM.
\end{IEEEbiography}

\vspace{-80pt}

\begin{IEEEbiography}
[{\includegraphics[width=0.8in,clip,keepaspectratio]{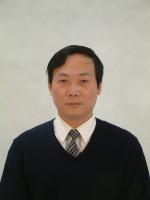}}]
{Jiankun Hu} is a Professor at the School of Engineering and IT, University of New South Wales (UNSW) Canberra (also named UNSW at the Australian Defence Force Academy (UNSW@ADFA), Canberra, Australia).
He is the invited expert of Australia Attorney-Generals Office assisting the draft of Australia National Identity Management Policy.
Prof. Hu has served at the Panel of Mathematics, Information and Computing Sciences (MIC), ARC ERA (The Excellence in Research for Australia) Evaluation Committee 2012.
His research interest is in the field of cyber security covering intrusion detection, sensor key management, and biometrics authentication.
He has many publications in top venues including IEEE Transactions on Pattern Analysis and Machine Intelligence, IEEE Transactions on Computers, IEEE Transactions on Parallel and Distributed Systems (TPDS), IEEE Transactions on Information Forensics \& Security (TIFS), Pattern Recognition, and IEEE Transactions on Industrial Informatics.
He is the associate editor of the IEEE Transactions on Information Forensics and Security.
\end{IEEEbiography}

\end{document}